\documentstyle[preprint,aps,prbbib,eqsecnum,epsf]{revtex}
\tighten
\begin{document}

\preprint{}

\draft

\title{Superconductivity in the two dimensional Hubbard Model}

\author{J. Beenen and D.M. Edwards}

\address{Department of Mathematics, Imperial College, 
London SW7 2BZ, UK}

\date{24 february 1995}

\maketitle

\begin{abstract}
Quasiparticle bands of the two-dimensional Hubbard model are 
calculated using 
the Roth two-pole approximation 
to the one particle Green's function.
Excellent agreement is obtained with recent Monte Carlo 
calculations, including an  
anomalous volume of the Fermi surface near half-filling, which can 
possibly be explained 
in terms of a breakdown of Fermi liquid theory. 
The calculated bands are very flat around the $(\pi,0)$ points of 
the Brillouin zone 
in agreement with photoemission measurements of cuprate 
superconductors. With doping there is a shift in spectral weight 
from the upper band to the lower band.
The Roth method is extended to deal 
with superconductivity within a four-pole approximation allowing 
electron-hole mixing. 
It is shown that 
triplet $p$-wave pairing never occurs.
Singlet $d_{x^2-y^2}$-wave pairing is strongly favoured and optimal 
doping occurs 
when the 
van Hove singularity, corresponding to the flat band part, lies at 
the Fermi level. 
Nearest neighbour antiferromagnetic 
correlations play an important role in flattening the bands near 
the Fermi level and in favouring  superconductivity.
However the mechanism for superconductivity is a local one, in
contrast to spin fluctuation exchange models. 
For reasonable values of the hopping 
parameter the transition temperature $T_c$ is in the range 10-100K.
The optimum doping $\delta_c$ lies between $0.14$ and $0.25$, 
depending on the 
ratio $U/t$. The gap equation has a BCS-like form and
$2\Delta_{max}/kT_c\simeq 4$.
\end {abstract}


\narrowtext

\section{Introduction}
In 1957 Bardeen, Cooper and Schrieffer~\cite{BCS} provided, 
at a stroke, the essentials of
a complete theory of the phenomenon of superconductivity 
as it was known then and for the
next thirty years. However since the discovery of high 
temperature superconductivity in
cuprate materials~\cite{bedmue} it has been generally, 
although not universally, believed
that a new mechanism is operating in these systems. It 
was established early on that one key
element of the BCS mechanism, electron pairing, 
remains.\cite{gough} 
However there is mounting
evidence that the symmetry of the pairs is
$d_{x^2-y^2}$~\cite{four,five,six,seven,eight} rather 
than $s$-wave and
this suggests an electronic mechanism for pairing rather 
than the original BCS phonon-mediated
one.

The common element in all the cuprate superconductors 
is the $CuO_2$ plane in which the $Cu$
atoms form a square lattice with an $O$ atom at the midpoint 
of each pair of nearest-neighbour
$Cu$ atoms. In the simplest case of $La_2CuO_4=(LaO)_2CuO_2$, 
with the assumption of
$La^{3+}$ and $O^{2-}$ ions, the $Cu$ charge is 2+ corresponding 
to a $3d^9$ configuration. In
the presence of crystal field splitting one expects doubly 
occupied $d_{xy}$, $d_{yz}$, $d_{zx}$ and
$d_{3z^2-r^2}$ orbitals and a singly-occupied $d_{x^2-y^2}$ 
orbital. In the absence of
interaction between the electrons the undoped system $La_2CuO_4$ 
would therefore be a metal
whereas it is observed to be an antiferromagnetic insulator. 
This demonstrates the importance
of strong repulsive Coulomb interaction on the $Cu$ site which 
tends to localize the $d$
electrons and produces a Mott insulator. In the doped system 
$La_{2-x}Sr_xCuO_4$, where
$La^{3+}$ ions are replaced by $Sr^{2+}$, the nominal 
occupation of the $x^2-y^2$ $Cu$ orbital
is reduced to $1-x$ and with $x\sim 0.15$ the system is 
metallic and superconducting with
$T_c\sim 35$K.

Anderson~\cite{pwa-sci} was the first to propose that the 
essence of high temperature
superconductivity is contained in the two dimensional (2D) 
square lattice Hubbard 
model~\cite{hubbardI} with
repulsive on-site interaction $U$. The atomic orbital in the model 
may be regarded as a $Cu$
$d_{x^2-y^2}$ orbital hybridized with $O$ $p_x$ and $p_y$ orbitals 
on neighbouring sites.\cite{zhangrice,esksaw} In
his recent work Anderson~\cite{anderson} attributes the existence 
of superconductivity to
electron transfer \underline{between} $CuO_2$ planes. He proposes 
that electrons in the
$CuO_2$ plane separate into uncharged spinons and charged holons, 
as they do in the
\underline{one} dimensional Hubbard model, where the electrons form 
a 
Luttinger liquid. This
spin-charge separation inhibits the transfer of unpaired electrons 
between planes, but in
Anderson's theory this constraint is removed upon pairing. The 
resultant decrease in total
kinetic energy favours pairing and drives superconductivity. A 
difficulty with Anderson's
theory is the absence of convincing evidence for spin-charge 
separation in two dimensions. 

From the experimental point
of view it is true that the critical temperature $T_c$ tends to 
increase with the number of 
adjacent $CuO_2$ planes, although $Tl_2 Ba_2 CuO_6$ with rather 
isolated $CuO_2$ planes has
$T_c=85$K.\cite{Tcsingleplane} Some interplanar interaction is 
presumably essential to
stabilize superconductivity at finite $T$ even when the primary 
mechanism, which sets the
scale of the resultant $T_c$, operates within a plane. An analogous 
situation is a quasi-2D
ferromagnet without magnetic anisotropy in which the scale of the 
Curie temperature is set by
in-plane exchange interaction, although weak interplanar exchange  
is required to stabilize
ferromagnetism at finite $T$.

An in-plane mechanism which is advocated by Pines and co-workers,
\cite{pines}
Scalapino,\cite{scalapino} and Moriya\cite{moriya} is based on 
electron pairing due to
exchange of antiferromagnetic spin fluctuations. This type 
interaction leads inevitably to
$d_{x^2-y^2}$ pairing and much of the experimental evidence cited 
in its favour is evidence
for $d_{x^2-y^2}$ pairing rather than for the spin fluctuation 
mechanism itself. In this
paper we show how $d_{x^2-y^2}$ pairing arises in the 2D Hubbard 
model in a way which is not
obviously related to antiferromagnetic spin fluctuations, although 
antiferromagnetic
correlations between neighbouring sites are clearly present. The 
spin fluctuation mechanism
has been strongly criticized by Anderson,\cite{pwa-phystoday} who 
emphasizes the heuristic
nature of Pines's model.

The published ab initio Monte Carlo calculations on the 2D Hubbard 
model have not produced
any evidence of superconductivity.\cite{vdL-MC} However for the 
small finite size systems
considered the electron density cannot be varied continuously and 
it is possible that the
favourable case of optimal doping is missed. Very recently Husslein 
{\it et
al.}\cite{husslein} have addressed this question by implementing 
projector Monte Carlo
calculations for the 2D Hubbard model with nearest and next-nearest 
neighbour hopping
parameters $t,t'$. This work was motivated by the `van Hove 
scenario'~\cite{newns} in which,
given a suitable interaction between electrons, a high $T_c$ may 
emerge from the large density
of states at the van Hove singularity associated with saddle-points 
in the band structure.
To model a particular system $t'/t$ is tuned so that for optimal 
doping, i.e. with a
convenient electron density close to that with the highest observed 
$T_c$, the singularity
occurs at the Fermi energy. In earlier work~\cite{newns} on the 
`van Hove scenario'
explicitly attractive electron interactions, such as the phonon 
mediated one, were considered
and consequently the pairing was $s$-wave. The remarkable result 
of the new Monte Carlo
calculations~\cite{husslein} is that $d_{x^2-y^2}$ pairing emerges 
in the repulsive $U$
Hubbard model, but only when, for a given $t'/t$, the doping level 
is tuned to within $\pm 5$\% of the optimum one. 
It is clear why superconductivity was 
missed in earlier Monte Carlo
calculations. The new  Monte Carlo results provide very satisfying 
confirmation of some of
the results reported in this paper, which are obtained by a more 
analytical Green's function
method. We will show how a combination of on-site electron 
repulsion and the band structure
saddle-point leads to $d_{x^2-y^2}$ pairing 
and a fairly high $T_c$.
Our approach gives a
unified theory of quasiparticles in the normal and superconducting 
states. A slight surprise
about the work of Husslein {\it {\it et al}.}~\cite{husslein} is 
that their modest interaction
strength $U=2t$ leads to sufficient correlation to give the effect.

We use a Green's function decoupling scheme originally due to 
Linderberg and
\"{O}hrn\cite{lind-ohrn} and first applied to calculations on the 
Hubbard model by
Roth.\cite{roth} The formalism is reviewed in 
Sec.~\ref{section:form} and
in Sec.~\ref{section:normalstate} it is applied to the normal 
paramagnetic state of the 2D Hubbard
model. It is shown that the Roth two-pole approximation gives 
excellent agreement with the
quasiparticle dispersion curves found in recent Quantum Monte 
Carlo results by Bulut {\it et
al.}\cite{bulut}. These authors have shown that their results are 
consistent with recent
angular resolved photoemission measurements of the hole-doped 
cuprates. A natural extension
of the method to superconductivity, now using a four-pole 
approximation, is made in
Sec.~\ref{section:scstate}. We find a superconducting state
in which the gap is determined by a non-standard 
correlation function.
In Sec.~\ref{section:gap} a gap equation is derived in two ways,
yielding a lower and upper bound to the gap. It is shown that 
triplet 
pairing cannot occur, but that singlet
$d_{x^2-y^2}$ pairing is strongly favoured, with $T_c=10-100$K.
The critical temperature $T_c$ is strongly dependent on doping, 
with the optimum doping corresponding to the case where the 
saddle-point 
of the band structure is situated exactly at the Fermi level.
This links our work to the `van Hove scenario', although none
of its ideas were imposed beforehand, unlike work by other 
authors.\cite{dagotto} Our results arise from the Hubbard model
and the decoupling scheme only.

\section{The formalism}\label{section:form}
In this paper we wish to determine Green's functions of the Hubbard 
model~\cite{hubbardI}
with hamiltonian
\begin{equation}
\hat{H}=t\sum_{<i,j>}\sum_{\sigma}c_{i\sigma}^{\dagger}c_{j\sigma}+
U\sum_{i}n_{i\sigma}n_{i-\sigma}-\mu\sum_{i,\sigma} n_{i\sigma},
\end{equation}
including the chemical potential in a standard notation.
The equation of motion for a retarded Green's function
$\left<\left<\hat{A};\hat{B}\right>\right>$ takes the form 
\begin{equation}
\omega\left<\left<\hat{A};\hat{B}\right>\right>=\left<\left[\hat{A},
\hat{B}\right]_+\right>+\left<\left<\left[\hat{A},\hat{H}\right];
\hat{B}\right>\right>,\label{eq:EOM}
\end{equation}
where $<...>$ denotes the thermal average.
Note the introduction of a new Green's function 
$\left<\left<\left[\hat{A},\hat{H}\right];\hat{B}\right>\right>$,
which implies that an infinite set of equations needs to be solved.
A very satisfying way to decouple this set of equations is to 
introduce a set of operators $\{\hat{A}_n\}$, which are
believed to be the most relevant to describe the one particle 
excitations of the system of interest.
Formally this assumption is that
\begin{equation}
\left[\hat{A}_n,\hat{H}\right]=\sum_mK_{nm}\hat{A}_m. \label{eq:K}
\end{equation}

Here, for the application of the formalism on the 2D Hubbard
model in the thermodynamic limit for several parameter ratio's 
$U/t$, we follow Roth~\cite{roth} in
choosing two operators of Bloch type
\begin{equation}
\hat{A}_{1\vec{k}\sigma}=c_{\vec{k}\sigma}=\frac{1}
{\sqrt{L}}\sum_{i}e^{i\vec{k}.\vec{R_{i}}}c_{i\sigma},
\label{eq:A1}\end{equation}
\begin{equation}
\hat{A}_{2\vec{k}\sigma}=d_{\vec{k}\sigma}=\frac{1}{\sqrt{L}}
\sum_{i}e^{i\vec{k}.\vec{R_{i}}}n_{i-\sigma}c_{i\sigma}
\label{eq:A2}\end{equation}
to describe the normal, spatially uniform, state of the system. 
Here $L$ is the
number of lattice sites in the system.
The restriction to spatial uniformity allows discussion of 
ferromagnetic states,\cite{roth,BEfer}
but additional operators are necessary to describe 
antiferromagnetic states.\cite{eapen,lind-hew}\\
To describe superconducting states, we shall also introduce 
additional `hole'
operators to supplement the `electron' operators 
(\ref{eq:A1}),(\ref{eq:A2}):
\begin{equation}
\hat{A}_{3\vec{k}\sigma}=c_{-\vec{k}-\sigma}^{\dagger}=\frac{1}
{\sqrt{L}}\sum_{i}e^{i\vec{k}.\vec{R_{i}}}c_{i-\sigma}^{\dagger},
\label{eq:A3}\end{equation}
\begin{equation}
\hat{A}_{4\vec{k}\sigma}=d_{-\vec{k}-\sigma}^{\dagger}=\frac{1}
{\sqrt{L}}\sum_{i}e^{i\vec{k}.\vec{R_{i}}}n_{i\sigma}
c_{i-\sigma}^{\dagger}.
\label{eq:A4}\end{equation}

The coefficients $K_{nm}$ in Eq. (\ref{eq:K}) are determined by
anticommuting both sides of Eq. (\ref{eq:K}) with each element 
of the
operator set $\{\hat{A}_n\}$ and then taking the thermal average. 
This can be
written in matrix notation as
\begin{equation}
\bf{E}=\bf{K}\bf{N},\label{eq:EKN}
\end{equation}
where the energy and normalization matrices, $\bf{E}$ and $\bf{N}$, 
are given by
\begin{equation}
E_{nm}=\left<\left[\left[\hat{A}_n,\hat{H}\right],
\hat{A}_m^\dagger\right]_
{+}\right>,\label{eq:E}
\end{equation}
\begin{equation}
N_{nm}=\left<\left[\hat{A}_n,\hat{A}_m^\dagger\right]_{+}\right>.
\label{eq:N}
\end{equation}
Combining Eqs. (\ref{eq:EOM}), (\ref{eq:K}), and (\ref{eq:EKN})-
(\ref{eq:N}) we obtain
\begin{equation}
\left<\left<\hat{A}_n;\hat{B}\right>\right>=\sum_m \tilde{G}_{nm}\;
\left<\left[\hat{A}_m, 
\hat{B}\right]_+ \right>,\label{eq:GtildeAB}
\end{equation}
where $\tilde{\bf{G}}$ is given by
\begin{equation} 
\tilde{\bf{G}}=\bf{N}({\omega}\bf{N}-\bf{E})^{-1}. \label{eq:Gtilde}
\end{equation}
If $\hat{B}=\hat{A}_m^\dagger$ we obtain for the Green's function 
matrix, whose elements are given
 by
\begin{equation}
G_{nm}(\omega)=\left<\left<\hat{A}_{n};\hat{A}_{m}^\dagger\right>
\right>\label{eq:Gij},
\end{equation}
a solution in terms of the matrices $\bf{E}$ and $\bf{N}$:
\begin{equation}
\bf{G}(\omega)=\bf{N}({\omega}\bf{N}-\bf{E})^{-1}\bf{N}.\label{eq:G}
\end{equation}
This decoupling procedure was first proposed by Linderberg and 
\"{O}hrn.\cite{lind-ohrn}
Soon after Roth~\cite{roth} applied this procedure to study 
ferromagnetism in the 3D infinite $U$
Hubbard model in the thermodynamic limit.
It can be shown that the formalism is essentially equivalent to the 
Mori-Zwanzig
projection technique~\cite{mo-zwa,fulde,mehlig} and strongly 
related to
moment methods~\cite{lonke,bowen} based on the  
assumption that the spectral function 
is a finite sum of weighted
delta functions.\cite{har-lan,nolting}

The  matrix elements $E_{nm}$ and $N_{nm}$ involve correlation 
functions which should
be determined self-consistently from the calculated Green's 
functions as far as
possible. Sometimes however, as discussed in 
Sec.~\ref{section:normalstate}, one must introduce
further Green's functions. The standard relationship between 
correlation
function and Green's function may be written as
\begin{equation}
\left<\hat{B}\hat{A}\right>=\frac{1}{2\pi i}
\oint
f(\omega)\left<\left<\hat{A};\hat{B}\right>\right>_{\omega}
d\omega,\label{eq:Fw}
\end{equation}
where $f(\omega)$ is the Fermi
function $f(\omega)=(e^{\beta\omega}+1)^{-1}$ and the contour 
encircles the real
axis without enclosing any poles of $f(\omega)$.
The chemical potential $\mu$ is determined by the condition
\begin{equation}
n=L^{-1}\sum_{\vec{k}} \left<c_{\vec{k}\sigma}^\dagger 
c_{\vec{k}\sigma}\right>=
\frac{1}{2\pi i L}\sum_{\vec{k}} \oint f(\omega) 
G_{11\vec{k}}(\omega)d \omega,
\end{equation}
where $n=\left< n_{i\sigma}\right>$ is  the average site occupation 
per spin.

\section{The normal state}\label{section:normalstate}
For the normal paramagnetic state of the Hubbard model, within the 
2-operator
set $\{\hat{A}_{1\vec{k}\sigma}, \hat{A}_{2\vec{k}\sigma}\}$, the 
energy and
normalization matrices $\bf{E}_2$, $\bf{N}_2$ are given 
by~\cite{roth}
\begin{equation}
\bf{E}_2=
\left[
\begin{array}{cc}
\epsilon_{\vec{k}}+Un-\mu&(\epsilon_{\vec{k}}-\mu+U)n\\
(\epsilon_{\vec{k}}-\mu+U)n&(U-\mu)n+\epsilon_{\vec{k}}n^2
+n(1-n)W_{\vec{k}}
\end{array}\right]\label{eq:E2}
\end{equation}
\begin{equation}
\bf{N}_2=\left[\begin{array}{cc}1&n\\
n&n\end{array}\right].\label{eq:N2}
\end{equation}
The Green's function matrix is readily found using Eq. (\ref{eq:G}).
Here $\epsilon_{\vec{k}}$ is the unperturbed band energy
\begin{equation}
\epsilon_{\vec{k}}=t\sum_{<j>_i}e^{i\vec{k}.(\vec{R}_j-\vec{R}_i)}
\label{eq:eps}
\end{equation}
and $W_{\vec{k}}$ may be written as:
\begin{equation}
n(1-n)W_{\vec{k}}=w_0+w_1\epsilon_{\vec{k}}\;,\label{eq:W}
\end{equation}
with 
\begin{equation}
w_0= -\sum_{<j>_i} t\left<  c_{i\sigma}^\dagger  c_{j\sigma}
(1- n_{i-\sigma}- n_{j-\sigma})\right>,\label{eq:w0}
\end{equation}
\begin{eqnarray}
w_1&=&
\frac{1}{4}\left(\left<N_jN_i\right>-\left<N_j\right>
\left<N_i\right>\right)\nonumber\\
&&+\left<\vec{S}_j.\vec{S}_i\right>-\left< c_{j\sigma}^\dagger 
c_{j-\sigma}^\dagger c_{i-\sigma} c_{i\sigma}\right>.
\label{eq:w1}\end{eqnarray}
Here $N_j= n_{j\sigma}+ n_{j-\sigma}$ and $\vec{S}_j$ are the total 
number operator and
the spin operator respectively for site $j$.
Sites $j$ and $i$ are nearest neighbours as indicated by the $j$ 
summation $<j>_i$ in
Eqs. (\ref{eq:eps}) and (\ref{eq:w0}). For the 2D Hubbard
model on a square lattice the unperturbed band energy is given by 
\begin{equation}
\epsilon_{\vec{k}}= 2t\;(\cos k_xa+\cos k_ya) \label{eq:eps2}
\end{equation}
with $a$ the lattice constant.
To model the $CuO_2$ plane we take $t<0$ so that the bottom of the 
band is at
the $\Gamma$-point $k_x=k_y=0$.
The occupation number $n$ and the correlation functions in $w_0$ 
may be
calculated from Green's functions $G_{11}$ and $G_{12}$ by means of 
Eq. (\ref{eq:Fw}).
The correlation functions in $w_1$ cannot be determined directly in 
this way. A
natural way to determine the density and spin correlation functions 
in $w_1$
would be to extend the formalism to deal consistently with the 
corresponding
two-particle Green's functions. This would have the advantage of 
yielding the
spin dynamics of the system and work along these lines is in 
progress. Here,
however, we follow Roth's original procedure and introduce extra 
operators 
$\hat{B}_i$. Correlations of the form 
$\left<\hat{A}_1\hat{B}_i\right>$ can
then be calculated within the decoupling~(\ref{eq:K}) by using Eqs. 
(\ref{eq:GtildeAB}) and (\ref{eq:Fw}). The details are given in 
Roth's 
paper.\cite{roth}

\begin{figure}
\begin{center}
\leavevmode
\epsfxsize=3.2in
\epsffile{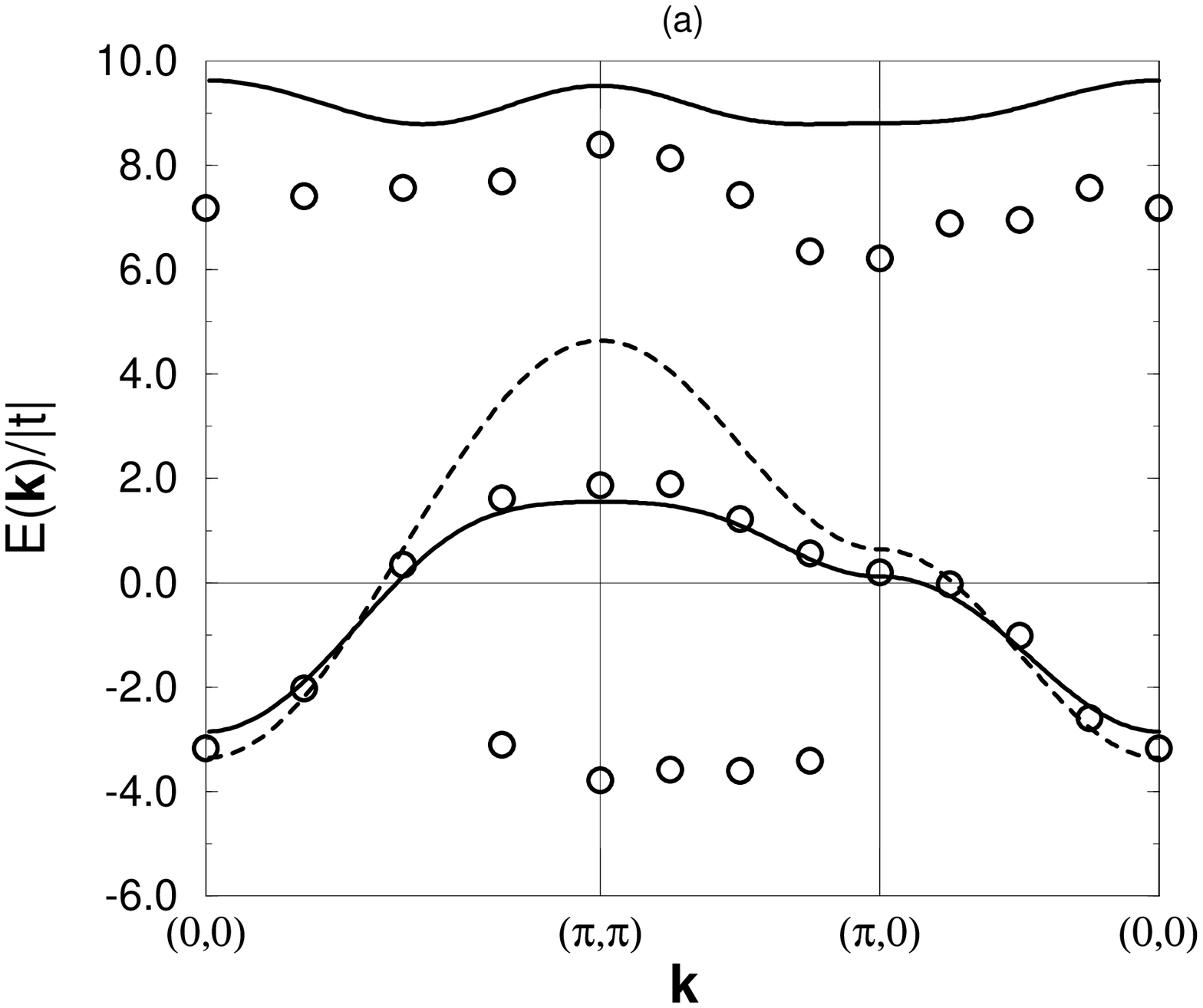}
\leavevmode
\epsfxsize=3.2in
\epsffile{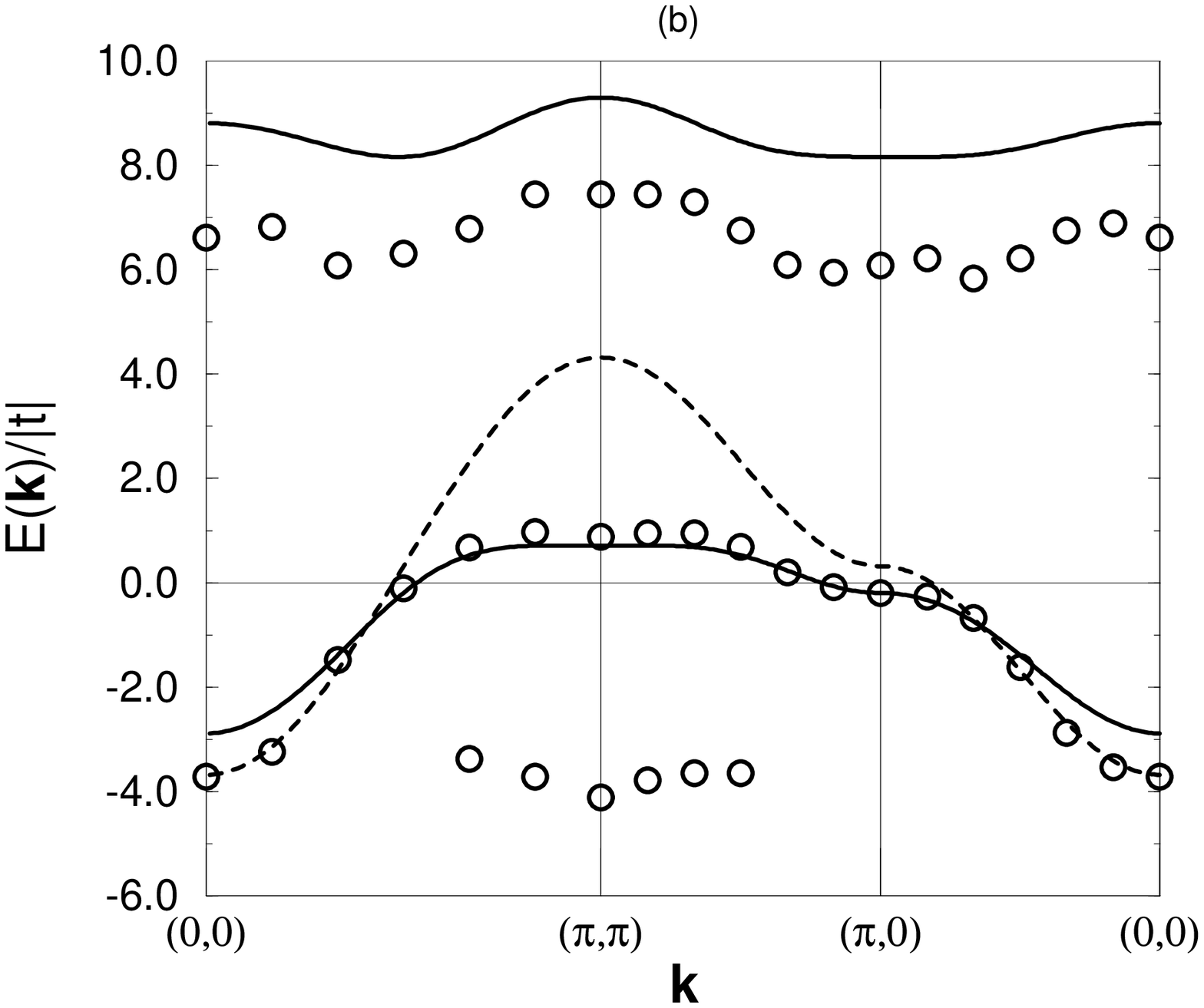}
\end{center}

\begin{center}
\leavevmode
\epsfxsize=3.2in
\epsffile{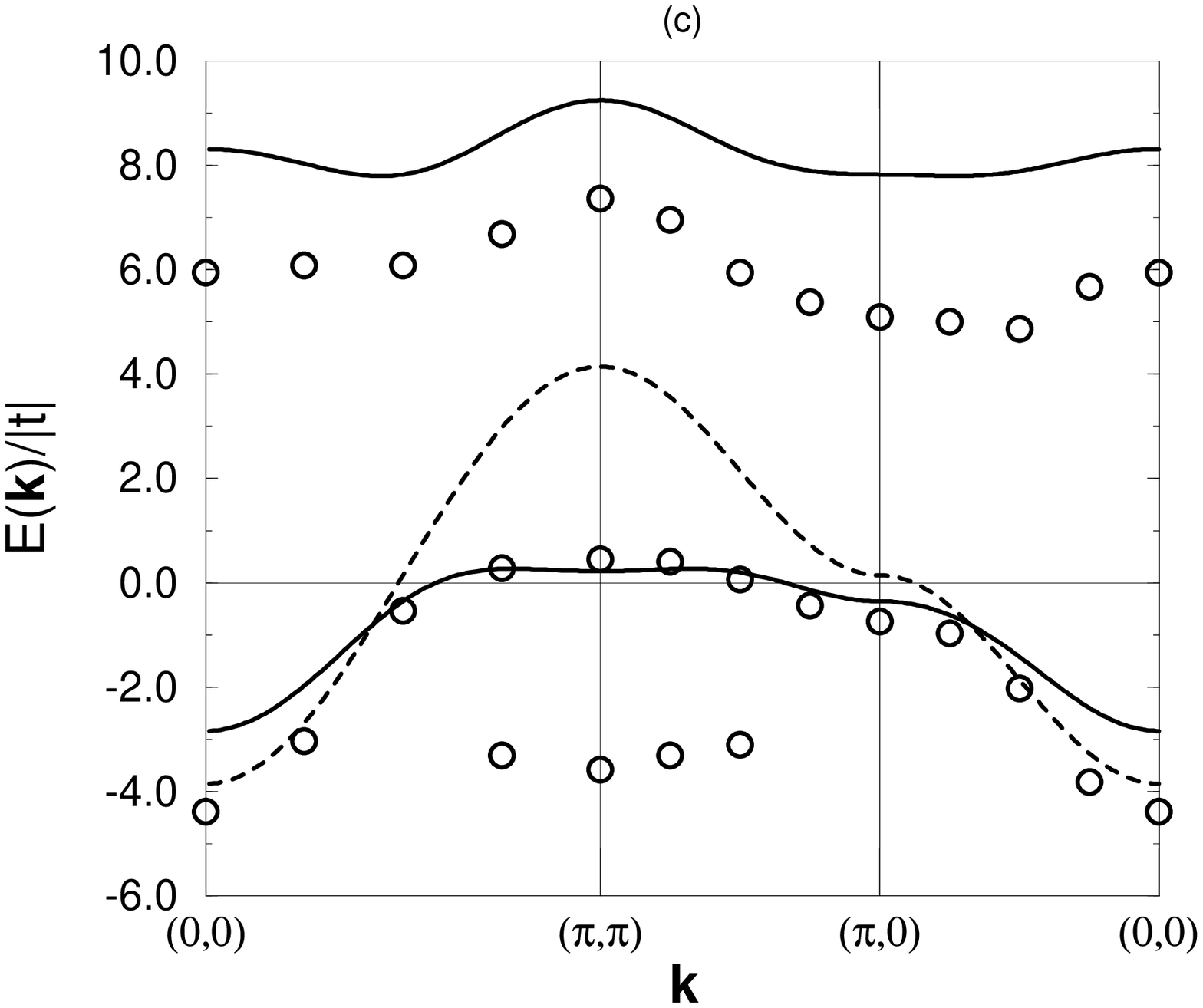}
\leavevmode
\epsfxsize=3.2in
\epsffile{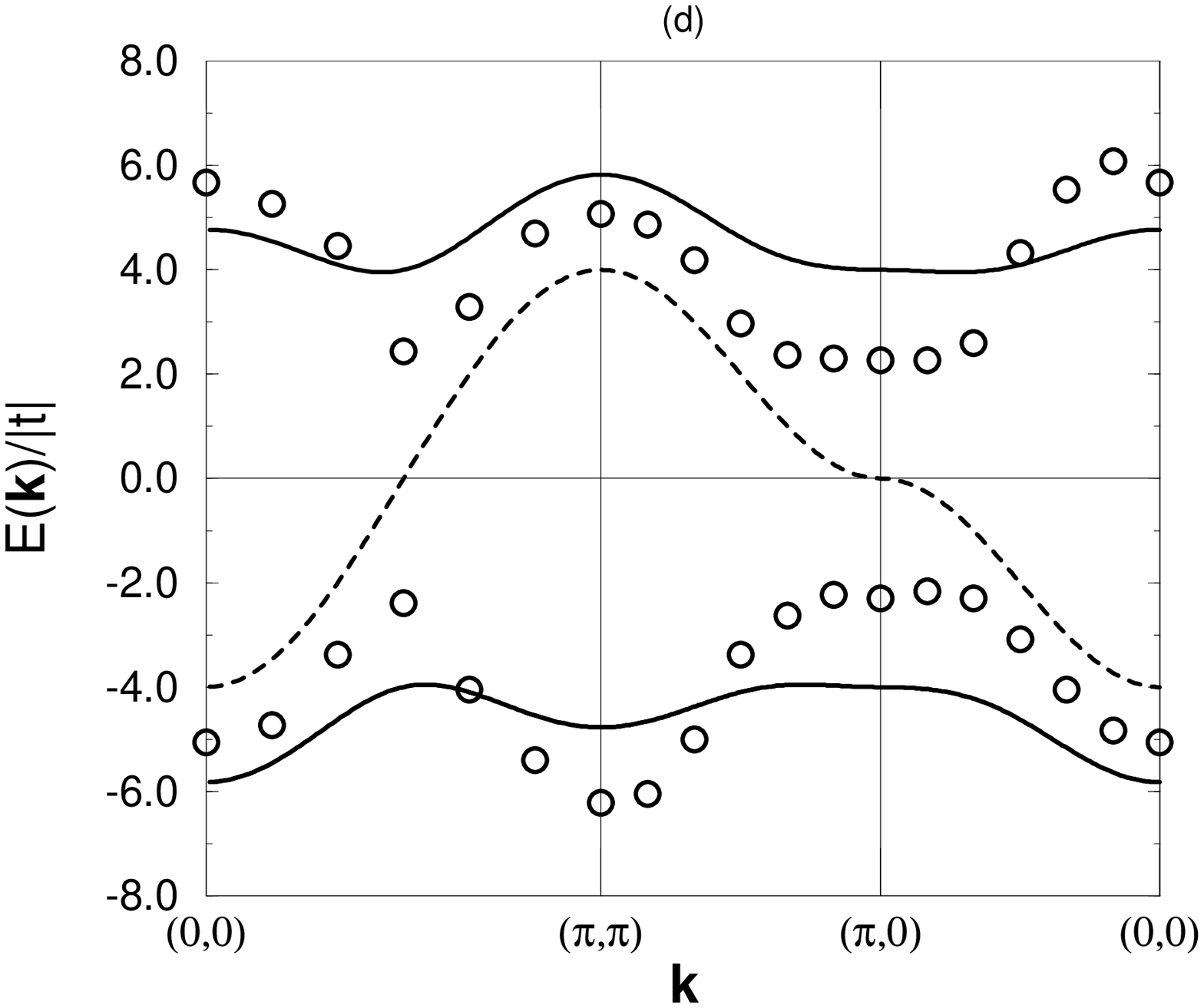}
\end{center}

\caption{The quasiparticle bands along the triangle
  \protect($0,0$) - $(\pi,\pi)$ - $(\pi,0)$  for several 
occupations $<N>$.
The solid lines are the calculated ones for $U=8|t|$; 
the circles are the results from the
Quantum Monte Carlo calculations by Bulut \protect{\it et al}.
\protect\cite{bulut} for the same parameters. The
dashed line indicates the band in the  noninteracting ($U=0$) 
case for the given occupation.
(a) $<N>=0.75$; (b) $<N>=0.87$; (c) $<N>=0.94$; and (d) $<N>=1.0$ 
(half-filling).}
\label{figure:bands}
\end{figure}

\begin{figure}
\begin{center}
\leavevmode
\epsfxsize=6.5in
\epsffile{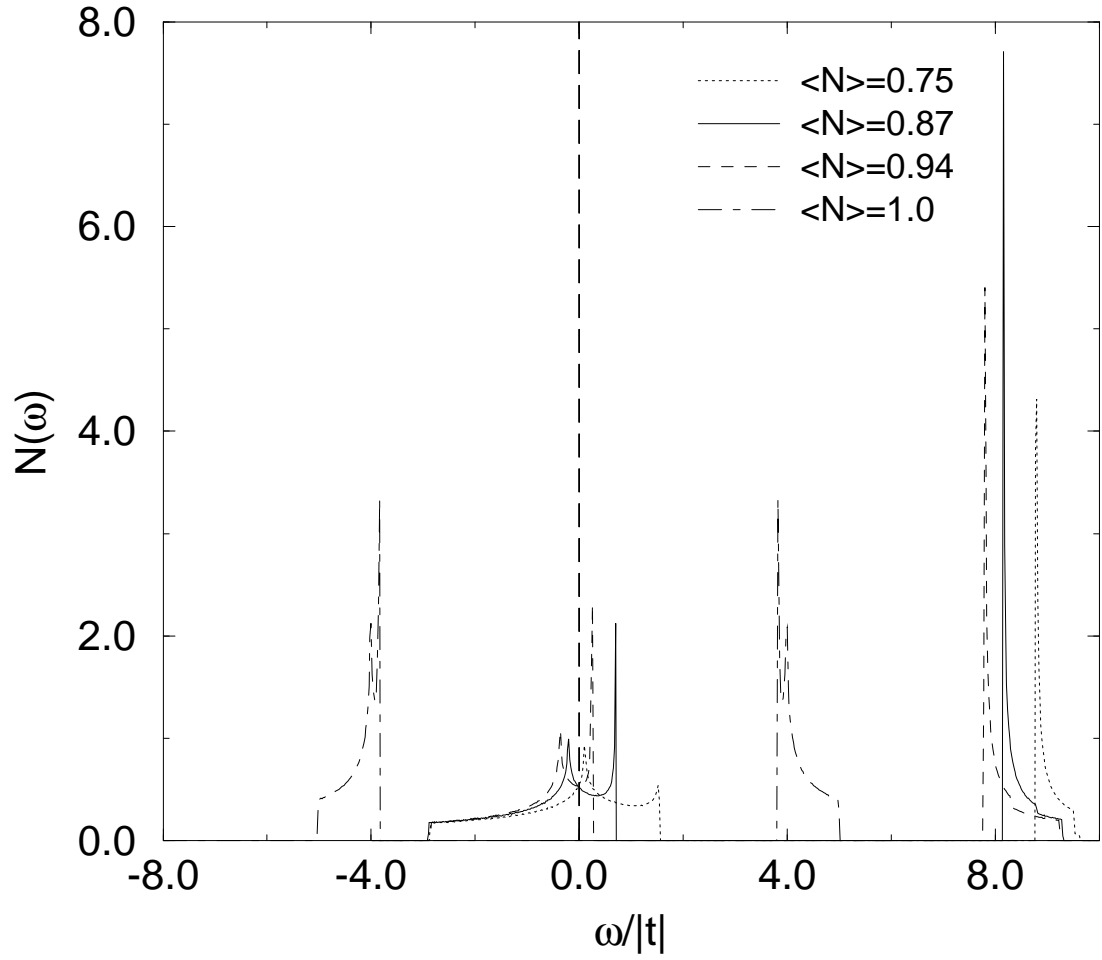}
\end{center}

\caption{The density of states for $U=8|t|$ and the occupations 
as in figure 1(a)-(d).}
\label{figure:dos}
\end{figure}

Thus to investigate the normal state of the 2D Hubbard model we
follow precisely the method Roth~\cite{roth} originally applied to 
the three dimensional cubic
lattice. The use of two operators, $\hat{A}_1$ and $\hat{A}_2$ 
corresponds to a
two-pole approximation to the Green's function.
The resulting Green's function therefore consists of two terms, 
each corresponding
to a quasiparticle band:
\begin{equation}
G_{11\vec k}(\omega)=\frac{\alpha_{1\vec k}}{\omega-\xi_{1\vec k}}
+\frac{\alpha_{2\vec k}}
{\omega-\xi_{2\vec k}}\label{eq:G11}
\end{equation}
with
\begin{equation}
\alpha_{1\vec k}=\frac{1}{2}+\frac{U(1-2n)-\epsilon_{\vec k}
+W_{\vec k}}{2X_{\vec k}}
\;\;\;\;;\;\;\;\;\alpha_{2\vec k}=1-\alpha_{1\vec k},
\end{equation}
\\
\begin{equation}
\xi_{1\vec k}=\frac{U+\epsilon_{\vec k}+W_{\vec k}-2\mu}{2} - 
\frac{X_{\vec k}}{2}\;\;\;\;
;\;\;\; \xi_{2\vec k}=\xi_{1\vec k} + X_{\vec k}, \label{eq:xi12}
\end{equation}
and
\begin{equation}
X_{\vec k}=\sqrt{(U-\epsilon_{\vec k}+W_{\vec k})^2
+4nU(\epsilon_{\vec k}-W_{\vec k})}.
\label{eq:X}\end{equation}
It has been shown that this Green's function conserves the first 
four moments 
of the spectral density.\cite{nolting}
The quasiparticle bands $\omega=\xi_{1\vec k}$ and $\omega=
\xi_{2\vec k}$ satisfy
$det\left(\omega\bf{N}_2-\bf{E}_2\right)=0$. 
They are plotted in Fig.~\ref{figure:bands}, for $U=8|t|$ and 
different site
occupations $\left<N\right>=2n$, along symmetry lines in the two 
dimensional
Brillouin zone. The quasiparticle energies are relative to the 
chemical
potential $\mu$ and comparison is made with the non-interacting band
$\epsilon_{\vec{k}}$, relative to the non-interacting chemical 
potential,
 and with very recent Quantum Monte Carlo results by Bulut
{\it et al}.\cite{bulut}. The agreement with the Monte Carlo 
results is remarkable and this is a
strong indication that the Roth two-pole approximation represents 
the Green's
function rather well. This is reasonable since the spectral 
functions of Bulut
{\it et al}.\cite{bulut} are dominated by two peaks which define 
the quasiparticle energies for
each $\vec{k}$. Near $\vec{k}=(\pi,\pi)$ there is also an 
additional broad bump
at lower energy, but its weight is very small. One striking 
feature is the
flattening at the top of the lower band around $\vec{k}=(\pi,\pi)$, 
extending to
the saddle-point at $\vec{k}=(\pi,0)$ and 
halfway to $\vec{k}=(0,0)$.
Very flat bands have also been observed near the $(\pi,0)$ point
in recent angular resolved photoemission (ARPES) 
experiments on hole doped cuprate superconductors such as 
$Bi_2Sr_2CaCu_2O_{8+\delta}$ (Bi 2212),\cite{ARPES Bi2212} 
$Bi_2Sr_2CuO_6 $ (Bi2201),\cite{ARPES Bi2201}
$YBa_2Cu_4O_8$ (YBCO 124),\cite{ARPES YBCO124,ARPES Y124} and 
$YBa_2Cu_3O_7$ (YBCO 123).\cite{ARPES YBCO124,ARPES YBCO123}
It is clear
from our calculations that this feature is associated to a large 
extent with
nearest neighbour antiferromagnetic correlations 
$\left<\vec{S}_i.\vec{S}_j\right><0$.
Because $\left<\vec{S}_i.\vec{S}_j\right>$ is by far the most 
important contributor
to $w_1$ for the occupations shown, $w_1$ will be negative, 
so that $W_{\vec{k}}$ 
decreases with increasing
$\epsilon_{\vec{k}}$. This leads to a narrowing and, near the top, 
a flattening of 
the lower band. $\left<\vec{S}_i.\vec{S}_j\right>$ increases with 
occupation, so the 
effect is pronounced when approaching half-filling.
The flattening near the top is also responsible for a clear gap 
between the upper and 
lower band, even for $U$ no larger than the unperturbed bandwidth. 
The densities of states calculated from the one-particle
Green's function are shown in Fig.~\ref{figure:dos}.

\begin{figure}
\begin{center}
\leavevmode
\epsfxsize=3.5in
\epsffile{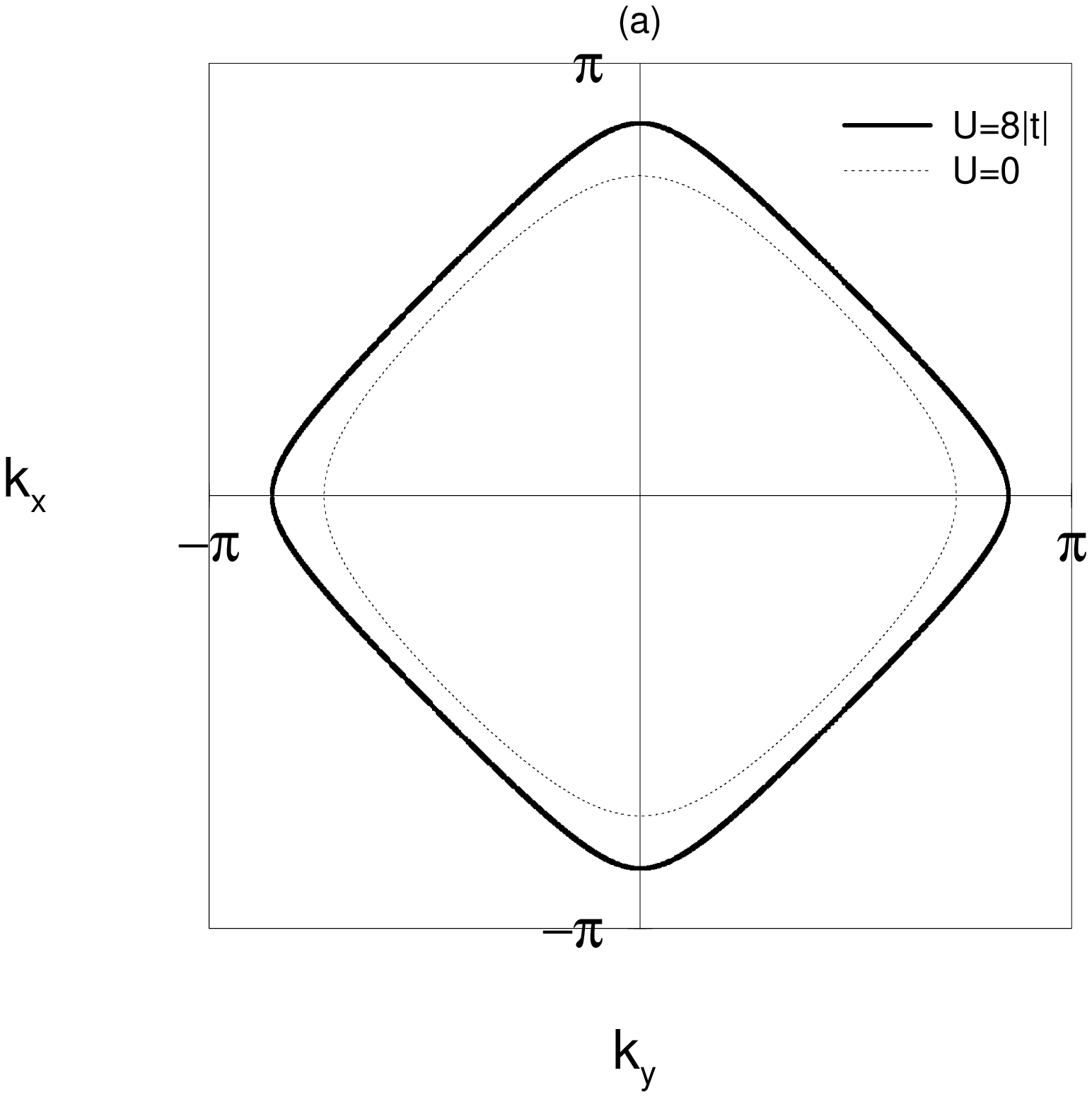}
\end{center}
\begin{center}
\leavevmode
\epsfxsize=3.5in
\epsffile{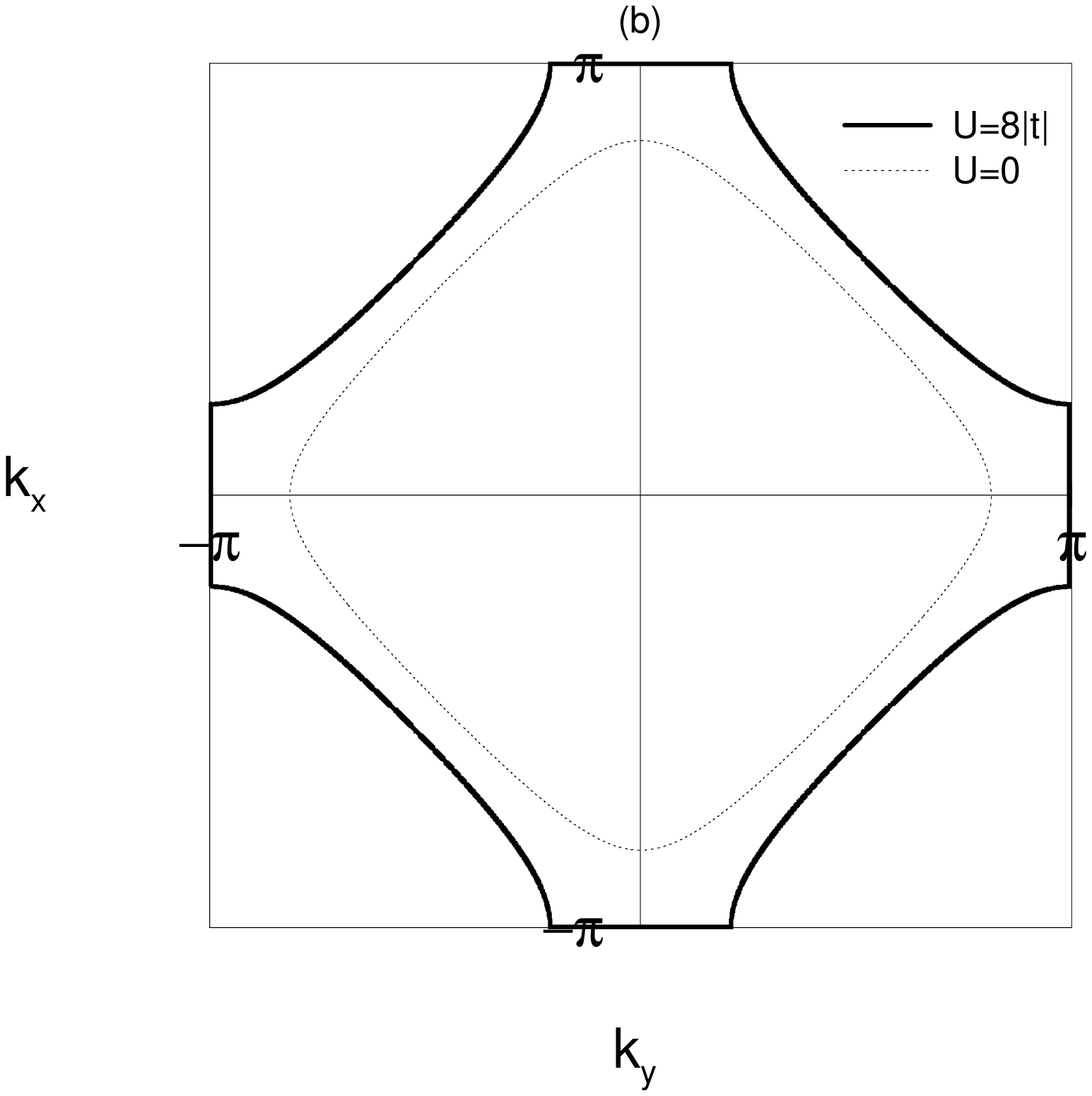}
\leavevmode
\epsfxsize=3.5in
\epsffile{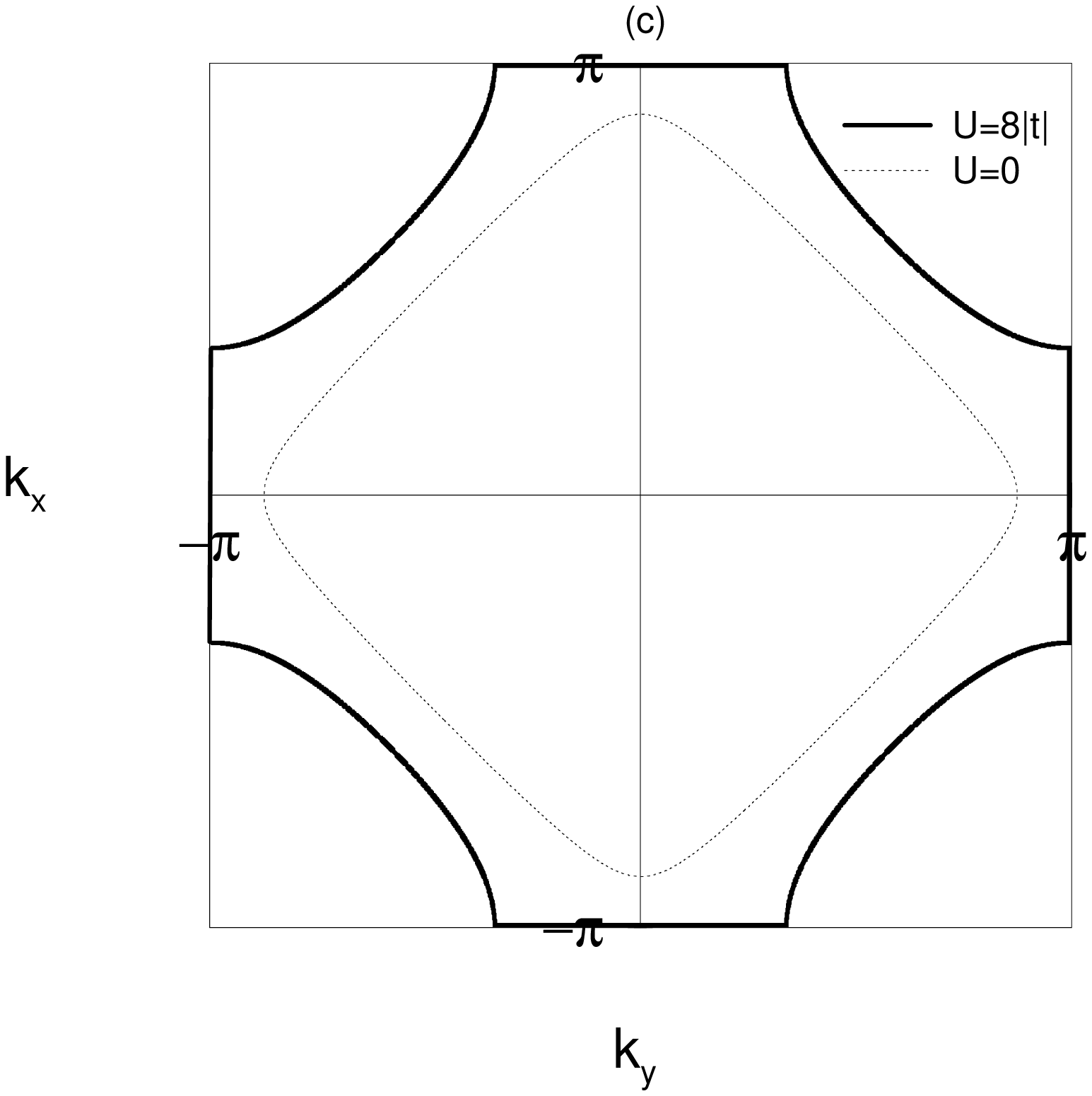}
\end{center}

\caption{The Fermi surface for $U=8|t|$ and the occupations as in 
figure 1(a)-(c). For comparison the Fermi
surfaces for the same occupations in the non-interacting 
case ($U=0$) are shown (dotted lines). }
\label{figure:FS}
\end{figure}

One feature that the Quantum Monte Carlo results
appears to share with our results is an anomalous Fermi surface 
volume for
$\left<N\right> \gtrsim 0.8$. In Fig.~\ref{figure:FS} we plot the 
Fermi surfaces
associated with our calculated bands and the non-interacting bands 
of Fig.~\ref{figure:bands}.
It is clear that the volume, or area in two dimensions, of the
calculated Fermi surface is always larger than the non-interacting 
one. Since
the Quantum Monte Carlo bands agree well with ours along symmetry 
lines, the
corresponding Fermi surfaces may be expected to be similar. 
Certainly for
$\left<N\right>=0.94$ it seems that the Quantum Monte Carlo data 
are definitely
incompatible with a Fermi surface of normal volume.
However, according to Luttinger's theorem~\cite{luttinger} the 
volume enclosed by the Fermi surface
of an interacting Fermi liquid is equal to that of the 
non-interacting system.
We would not want to conclude that there is a breakdown of 
Fermi liquid theory on the basis of
our own calculations, because, as will be discussed below, the 
two-pole approximation
will always lead to a Fermi surface enclosing an enlarged volume 
when only the
lower band is occupied.
But the Monte Carlo method in principle yields exact results for 
the model, so
the apparent anomalous Fermi surface volume must be taken seriously.
It must be remembered however that the Quantum Monte Carlo 
calculations~\cite{bulut} were
executed at a finite temperature, $T=0.5|t|$ in this case.
A possibility is that the enlarged Fermi surface volume is a 
temperature effect
and that the volume shrinks to its normal volume as 
$T\rightarrow 0$, although
this means that the form of the Fermi surface and the bands 
would change
considerably in the process.
In order that the chemical potential moves in the right 
direction with increasing
temperature, the Fermi level at $T=0$ would have to lie above the 
van Hove singularity
where the density of states has a negative gradient. Thus electron 
interaction would 
have produced a strong distortion of the Fermi surface which, 
although conserving
the enclosed volume, changes the topology. 
Another explanation might be that a transition from a Fermi liquid
to a non-Fermi liquid, of the type predicted by Edwards and Hertz,
\cite{EH} is
occurring as $\langle N \rangle$ increases through a critical value 
$N_c$. In
the Edwards-Hertz theory, based on a modified Hubbard alloy 
analogy, a $T=0$
phase transition occurs at a value of $N_c < 1$ when $U$ is 
larger than the
bandwidth. For $\langle N \rangle < N_c$ we have a normal Fermi 
liquid with a
Fermi surface of normal volume defined by a finite Migdal 
discontinuity
associated with quasiparticles of infinite lifetime. For 
$\langle N \rangle >
N_c$ there is no Migdal discontinuity since all quasiparticles 
have a finite
lifetime, attributed by Edwards and Hertz to very strong spin 
disorder
scattering. However clear quasiparticle peaks in the spectral 
functions still
exist, just as in the Monte Carlo data, and the points in 
$\vec k$ space
where they cross the chemical potential define a `Fermi surface'. 
It is found
\cite{marco} that for $\langle N \rangle > N_c$ the volume of this 
Fermi surface is enlarged
compared with the non-interacting one and spectral functions and 
quasiparticle
bands have been calculated for the 2D Hubbard model within the 
Edwards-Hertz
approach so that a detailed comparison with the Monte Carlo data 
can be made.
It is clear on general grounds that, if the
`Fermi surface' volume is indeed anomalous, one would not expect 
a Migdal
discontinuity with sharp quasiparticles because this is a vestige 
of the
non-interacting situation and would only occur at a surface of 
normal volume.

We now return to the present Roth approach which cannot describe 
a sharp transition
at a critical $N_c$ but which gives a smooth crossover from Fermi 
surfaces of almost normal
volume at smaller $\langle N \rangle$ to enlarged surfaces at
 $\langle N\rangle$ closer to 1.
The major defect of the two-hole approximation is that the 
quasiparticles, whose energies
appear to be given rather accurately, are not subject to 
any lifetime broadening. The origin
of the enlarged Fermi surface within this approximation is 
seen most clearly in the limit of
very large $U$ where the effect is most pronounced. For 
$U\rightarrow\infty$ the one-electron
Green's function $G_{11}$ is easily found from Eqs. 
(\ref{eq:G11})-(\ref{eq:X}) to take
the form 
\begin{eqnarray}
G_{11}&\simeq&\frac{1-n}{\omega-(1-n)\epsilon_{\vec{k}}-nW_{\vec{k}}
+\mu}\nonumber\\
&&+\frac{n}{\omega
-n \epsilon_{\vec{k}}-(1-n)W_{\vec{k}}-U+\mu}.
\end{eqnarray}
For $\langle N \rangle = 2n < 1$ only the lower band, associated 
with the
pole of the first term, is occupied. Since each state in this band 
has weight
$1-n$, whereas for a non-interacting case each state has weight 
$1$, the Fermi
surface volume is increased relative to the non-interacting one 
by a factor
$(1-n)^{-1}$. For example if $\langle N \rangle = \frac{2}{3}$, 
the enlargement
factor is $\frac{3}{2}$ and the Fermi surface volume corresponds 
to a half-filled
($\langle N \rangle = 1$) non-interacting situation. Since 
$W_{\vec k}$ depends
on ${\vec k}$ only through $\epsilon_{\vec k}$ , as is seen from Eq.
(\ref{eq:W}), it follows from Eq. (\ref{eq:xi12}) that a 
quasiparticle
constant energy surface enclosing a given volume coincides with the
non-interacting energy surface enclosing the same volume. 
Thus, with 
$U\rightarrow\infty$, for
$\langle N \rangle = \frac{2}{3}$ we have the square Fermi surface
$\epsilon_{\vec{k}}=0$, but for finite $U$, with a less 
pronounced enlargement
factor, the square Fermi surface occurs for larger 
$\langle N \rangle$. The
square Fermi surface plays an important role in the 
superconducting state
described in the following sections because, for given $U$, 
optimum doping
(highest $T_c$) occurs with the occupation $\langle N \rangle$ 
corresponding to
the square Fermi surface. This corresponds to having a van Hove 
singularity at the Fermi
level for some \underline{finite} doping and we have argued above 
that this must be the case
at $T=0$ to be consistent with the high temperature 
Monte Carlo data. Thus even if the
anomalous Fermi surface \underline{volume} of the Roth method 
is spurious we believe the
\underline{topology} of the Fermi surface is described quite well.

\begin{figure}
\begin{center}
\leavevmode
\epsfxsize=4.0in
\epsffile{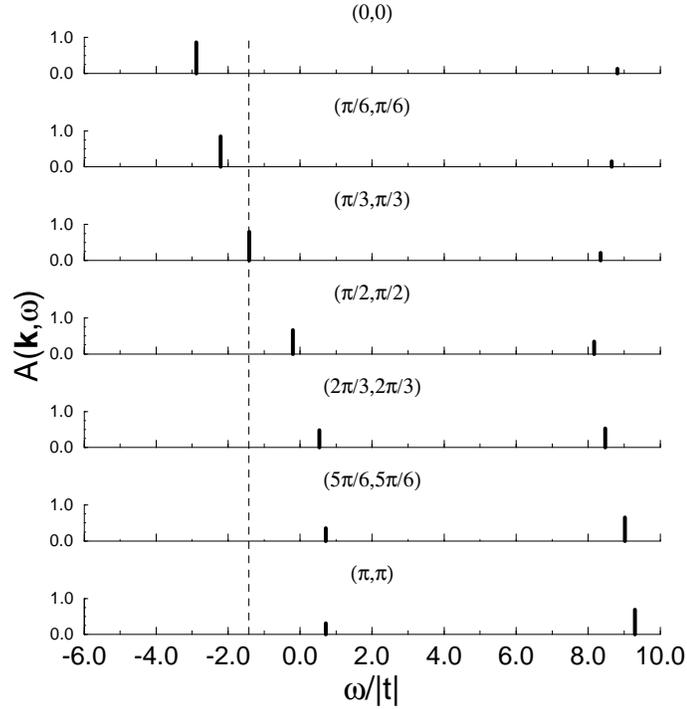}
\end{center}

\caption{The single particle spectral weight \protect 
$A(\vec{k},E)$, along the $k_x=k_y$ line,
 from our calculations. The height of the bars represents the
weight of the delta functions.
The weights of the delta functions compare well with the area's
under the peaks of Fig. 1(c) in Bulut \protect{\it et.al.},
Phys. Rev. B \protect{\bf 50}, 7215 (1994). ($<N>=0.87;\;U=8|t|$)}
\label{figure:weights}
\end{figure}

\begin{figure}
\begin{center}
\leavevmode
\epsfxsize=4.0in
\epsffile{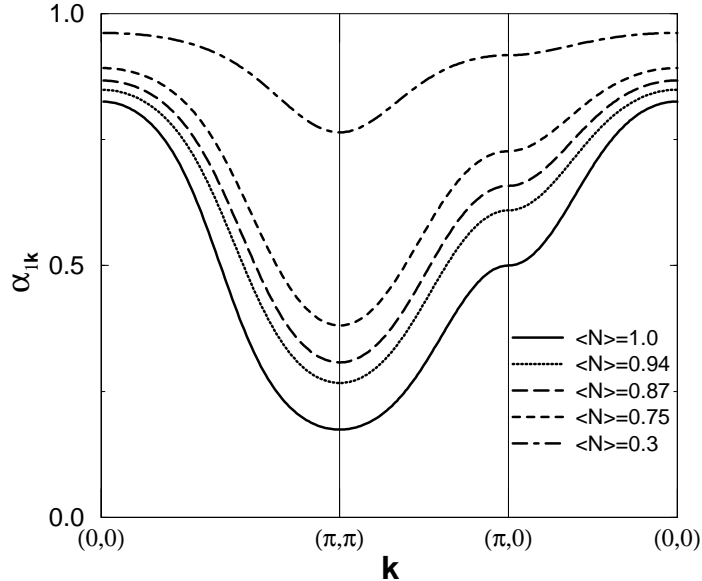}
\end{center}

\caption{The spectral weight $\alpha_{1\vec k}$ of the 
lower band for several occupations $<N>$. Note that the sum of the
spectral weights of the two bands is 1, so that this
 figure shows a weight transfer from the upper to the
lower band upon doping. ($U=8|t|$) }
\label{figure:weighttransfer}
\end{figure}

We wish to emphasize again the remarkable agreement shown in 
Fig.1 between the
quasiparticle bands calculated by the Roth two-pole approximation, 
using precisely her method
of evaluating various correlation functions which occur, and those 
given by the Monte Carlo
calculations of Bulut {\it et al}.\cite{bulut}.  Our calculations 
agree with the Monte Carlo data
not only for the location of the bands, but also for the spectral 
weight distribution over the
two bands. This is shown in Fig.~\ref{figure:weights} where 
we show, for several 
points along the $k_x=k_y$ line in the Brillouin zone, 
 our two delta functions with weights $\alpha_{1\vec k}$ 
and $\alpha_{2\vec k}$. These weights compare well with the areas 
under the peaks of the spectral functions found by 
Bulut {\it et al}.~\cite{bulut}   
While the spectral weight distribution 
is dispersionless for
$U\rightarrow  \infty$, for intermediate $U$ ($U=8|t|$) we find a 
strong dispersion.
We also find that there is a strong weight transfer upon doping 
from the upper
band 
$\omega=\xi_{2\vec k}$
to the lower band $\omega=\xi_{1\vec k}$, as shown in 
Fig.~\ref{figure:weighttransfer}.
This weight transfer is particularly strong around the ($\pi,\pi$) 
point for intermediate $U$.
A strong weight transfer from high energy  scale to low energy 
scale upon doping 
has been observed
in electron loss experiments~\cite{ELS} and optical spectroscopy 
experiments~\cite{OptSpec}
on high temperature cuprate superconductors. 
Applying strong-coupling perturbation theory,
Eskes {\it et al}.~\cite{eskes} find a similar weight transfer.

We may
mention here that the Roth procedure, again followed precisely, 
also gives a
good description of the occurence of ferromagnetism in the 2D 
Hubbard model.
The ferromagnetic region of the ($\langle N \rangle, t/U$) plane 
is found
to be \cite{BEfer} somewhat larger than the best variational 
estimates,\cite{vdL-E} but
the result is again remarkably good considering the subtlety of 
the energy
balance in this problem and the simplicity of the method. This 
contrasts strongly with the
Hubbard I approach,~\cite{hubbardI} which corresponds to setting 
$W_{\vec k}=0$.
Roth~\cite{roth} pointed out that the constant shift $w_0$ in 
$W_{\vec k}$ [see Eq.
(\ref{eq:W})] for the minority spin band is essential to stabilize 
the state of complete spin
allignment for large $U$ near half-filling. Recently Hewson and 
Wasserman~\cite{hewwas}
showed how an approach equivalent to the Roth one leads to the 
results of slave-boson mean
field theory in the Anderson impurity model; here again the shift 
in the impurity arising
from $w_0$ is essential to place the resonant state correctly near 
the Fermi level.
 These successes of the
present approach encourage us to consider another subtle problem,
superconductivity, in the next section. We shall see that an 
anomalous correlation function
which characterizes the superconducting state is generated 
automatically, just like the
correlation functions in $W_{\vec k}$.

\section{The superconducting state}\label{section:scstate}

To discuss superconductivity we need to mix electron and hole 
operators and
evaluate anomalous correlation functions in which particle number 
is not conserved.
In this context we note that the procedure described in 
Sec.~\ref{section:form},
 applied on the BCS reduced Hamiltonian and using the 
operator set $\{c_{\vec{k}\sigma},
c_{-\vec{k}-\sigma}^\dagger\}$ yields the whole BCS 
formalism~\cite{BCS} 
in Nambu-Gor'kov form.\cite{nam-gor}
As indicated in Sec.~\ref{section:form}, for the Hubbard 
model we use the four operators
(\ref{eq:A1})-(\ref{eq:A4}) and hence obtain a four-pole 
approximation to the
Green's functions. The  4x4 energy and normalization matrices 
$\bf{E}_4$ and
$\bf{N}_4$ may each be partitioned into four 2x2 matrices. The 
upper left
2x2 matrices are just $\bf{E}_2$ and $\bf{N}_2$, given by Eqs.
(\ref{eq:E2}) and (\ref{eq:N2}), and the lower right 2x2 matrices 
are easily
shown to be $-\bf{E}_2$ and $+\bf{N}_2$. The elements of the 
off-diagonal blocks
involve anomalous correlation functions and are as follows: 
\begin{equation}
N_{13}=N_{24}=N_{31}=N_{42}=0,
\end{equation}
\begin{equation}
N_{14}=-N_{23}=N_{41}^*=-N_{32}^*=\left< c_{i\sigma} 
c_{i-\sigma}\right>,
\end{equation}
\begin{equation}
E_{13}=E_{31}^*=U\left< c_{i-\sigma} c_{i\sigma}\right>,
\end{equation}
\begin{equation}
E_{14}=E_{32}^*=(\epsilon_{\vec{k}}-\mu)\left< c_{i\sigma} 
c_{i-\sigma}\right>,
\end{equation}
\begin{eqnarray}
E_{23}=E_{41}^*&=&(U-\epsilon_{\vec{k}}-\mu)\left< c_{i-\sigma} 
c_{i\sigma}\right>\nonumber\\
&&+\sum_{<l>_i}\;t\;\left< c_{i-\sigma} c_{l\sigma}+ c_{l-\sigma} 
c_{i\sigma}\right>,
\end{eqnarray}
\begin{eqnarray}
E_{24}&=&\sum_{<l>_i}
\;t\; e^{i\vec{k}.(\vec{R}_i-\vec{R}_l)}\left< n_{i-\sigma} 
c_{l\sigma} c_{l-\sigma}+ n_{l\sigma} c_{i\sigma} 
c_{i-\sigma}\right>\nonumber\\
&&+\sum_{<l>_i}\;t\;\left< n_{i\sigma} c_{i-\sigma} 
c_{l\sigma}- n_{i-\sigma} c_{i\sigma} c_{l-\sigma}\right>,
\end{eqnarray}
\begin{eqnarray}
E_{42}&=&\sum_{<l>_i}\;t\;e^{i\vec{k}.(\vec{R}_i-\vec{R}_l)}
\left< n_{i\sigma} c_{l-\sigma}^\dagger c_{l\sigma}^\dagger
+ n_{l-\sigma} c_{i-\sigma}^\dagger c_{i\sigma}^\dagger\right>
\nonumber\\
&&+\sum_{<l>_i}\;t\;\left< n_{i-\sigma} c_{i\sigma}^\dagger 
c_{l-\sigma}^\dagger- n_{i\sigma} c_{i-\sigma}^\dagger 
c_{l\sigma}^\dagger\right>.
\end{eqnarray}
These expressions simplify considerably when we introduce the 
symmetry
associated with either singlet or triplet pairing. The anomalous 
correlation
functions are matrix elements between states which differ by the 
addition of
one pair and the spin function associated 
with a pair is symmetric or
antisymmetric for triplet or singlet pairing respectively. It 
follows that if
all spin labels in the correlation function are reversed, i.e. 
$\sigma
\rightarrow -\sigma$, the sign changes in the singlet case but 
not in the triplet
case.

We first consider triplet pairing. Owing to the spin symmetry the 
second terms
in $E_{24}$ and $E_{42}$ vanish and $E_{13}=E_{14}=E_{23}=0$. 
Similarly all the
elements of the off-diagonal blocks in $\bf{N}_4$ vanish. 
Furthermore, denoting
the correlation function in the first term of $E_{24}$ by 
$\beta_{il}$, we have
\begin{eqnarray}
\beta_{il} &=& \langle n_{i-\sigma} c_{l \sigma} c_{l -\sigma} + 
n_{l \sigma} 
c_{i \sigma} c_{i -\sigma} \rangle  \nonumber\\
&=& \langle n_{i \sigma} c_{l -\sigma} c_{l \sigma} + n_{l -\sigma} 
c_{i -\sigma} c_{i \sigma} \rangle  \nonumber\\
&=& - \langle n_{l -\sigma} c_{i \sigma} c_{i -\sigma} + 
n_{i \sigma} 
c_{l \sigma} c_{l -\sigma} \rangle  \nonumber\\
&=& - \beta_{li}.\label{eq:betail}
\end{eqnarray}
Using this symmetry we write:
\begin{equation}
\beta_{il}=\left\{\begin{array}{llll} \beta&{\rm for}\;\;
\vec{R}_i-\vec{R}_l=(a,0)\\ -\beta&{\rm for}\;\;
\vec{R}_i-\vec{R}_l=(-a,0)\\
\pm\beta&{\rm for}\;\;
\vec{R}_i-\vec{R}_l=(0,a)\\ \mp\beta&{\rm for}\;\;
\vec{R}_i-\vec{R}_l=(0,-a).
\end{array} \right.. 
\end{equation}
Thus 
\begin{equation}
E_{24} = i \beta_{\vec{k}},\label{eq:E24p}
\end{equation}
where
\begin{equation}
\beta_{\vec k} = 2 t \beta (\sin k_xa \pm \sin k_ya).
\end{equation}
Similarly
\begin{equation}
E_{42} = - i \beta_{\vec k}^*.
\end{equation}
It will be seen shortly that $\beta_{\vec k}$ is essentially the gap
function and the nodes at $k_x = \pm k_y$ indicate $p$-wave 
symmetry. This is
as expected for triplet pairing. It is very interesting that the 
gap is
determined not by a simple pair wave function 
$\langle c_{i -\sigma} 
c_{l \sigma}
\rangle$, but by the more subtle correlation function $\beta_{il}$. 
This
quantity cannot be evaluated directly from our Green's functions. 
The simplest
way to evaluate it, and hence obtain a gap equation, is by a 
factorization
procedure described in Sec.~\ref{section:gap} for the case of 
singlet
$d$-wave pairing. We shall not pursue the $p$-wave case further 
here, because
this procedure, and others we have investigated, does not yield a 
non-zero
solution for the gap. The reason for this will be pointed out in
Sec.~\ref{section:gap}. We therefore find that $p$-wave 
superconductivity does
not occur in the 2D Hubbard model.
 
We now turn to singlet pairing and discuss particularly the case 
of $d$-wave
symmetry. We defer a discussion of $s$-wave symmetry to another 
occasion,
partly because we believe it is less likely and partly because the 
possibility
of on-site pairing $\langle c_{i -\sigma} c_{i \sigma} \rangle $ 
means that all
elements $E_{nm}$ in the off-diagonal blocks are non-zero. The 
$d$-wave case is
simpler because the $d$-symmetry of the pair wave function ensures 
that $\langle 
c_{i -\sigma} c_{i\sigma} \rangle = 0 $ and that the summations of 
the pair
wave function $\langle c_{i -\sigma} c_{l \sigma} \rangle $ and of 
$\langle 
 n_{i\sigma} c_{i-\sigma} c_{l\sigma}\rangle$ over sites $l$,
which are nearest neighbours of site $i$, give zero. The 
consistency of these
assertions can be checked by expressing these correlation 
functions in terms of
the resultant Green's function. Thus $E_{13}=E_{14}=E_{23}=0$ and, 
as in the
$p$-wave case, all elements of the off-diagonal blocks 
in $\bf{N}_4$ vanish.
In a similar way the second terms in $E_{24}$ and $E_{42}$ vanish 
and we are left
with just the first terms of $E_{24}$ and $E_{42}$, as in the 
$p$-wave case.
Denoting the correlation function in the first term of $E_{24}$ by
$\gamma_{il}$, we find, introducing the sign changes on spin 
reversal in the
singlet case, that the equation analogous to Eq. (\ref{eq:betail}) 
is:
\begin{equation} \gamma_{il} = \langle n_{i-\sigma} c_{l \sigma} 
c_{l -\sigma} +
n_{l \sigma}  c_{i \sigma} c_{i -\sigma} \rangle =
\gamma_{li}.\label{eq:gammail}  
\end{equation} 
Using the
$d$-wave symmetry we write 
\begin{equation} \gamma_{il} = \left\{\begin{array}{ll}\gamma&
{\rm for}\;\;{\vec R}_i - {\vec
R}_l = (\pm a,0) \\ - \gamma&{\rm for}\;\; 
{\vec R}_i - {\vec R}_l = (0,\pm
a)\end{array}\right. .
\end{equation} 
Hence
\begin{equation}
E_{24} = \gamma_{\vec k},\;\;\;\;\;   E_{42} = \gamma_{\vec k}^*, 
\label{eq:E24ds}
\end{equation}
where
\begin{equation}
\gamma_{\vec k} = t \sum_{\langle l \rangle_i} e^{i {\vec k}.
(\vec{R_l} -
\vec{R_i})} \gamma_{il} = g ( \cos k_xa - \cos k_ya)
\label{eq:gammak} \end{equation} with
\begin{equation}
g = 2 t \gamma.\label{eq:g}
\end{equation}
We call $\gamma_{\vec{k}}$ the gap-function and $g$ the 
gap-function amplitude.
The matrices $\bf{E}_4$ and $\bf{N}_4$ now take the partitioned form
\begin{equation}
\bf{E}_4=\left[\begin{array}{cc}\bf{E}_2&\begin{array}{cc}0&0\\
0&\gamma_{\vec{k}}\end{array}\\
\begin{array}{cc}0&0\\
0&\gamma_{\vec{k}}^*\end{array}&-\bf{E}_2\end{array}\right],
\label{eq:E4}
\end{equation}
\begin{equation}
\bf{N}_4=\left[\begin{array}{cc}\bf{N}_2&\emptyset\\
\emptyset&\bf{N}_2\end{array}\right]\;,\label{eq:N4}
\end{equation}
where $\bf{E}_2$ and $\bf{N}_2$ are given by Eqs. (\ref{eq:E2}) and
(\ref{eq:N2}).
We shall have a superconducting state if $\gamma_{\vec k} \ne 0$ 
and the
central problem we shall presently address is to find a gap 
equation satisfied
by $\gamma_{\vec k}$ and search for a non-zero solution. For the 
moment
we assume $\gamma_{\vec k} \ne 0$ and describe how to calculate the
Green's function matrix and discuss the general nature of the 
quasiparticle
bands.

\begin{figure}
\begin{center}
\leavevmode
\epsfxsize=4.2in
\epsffile{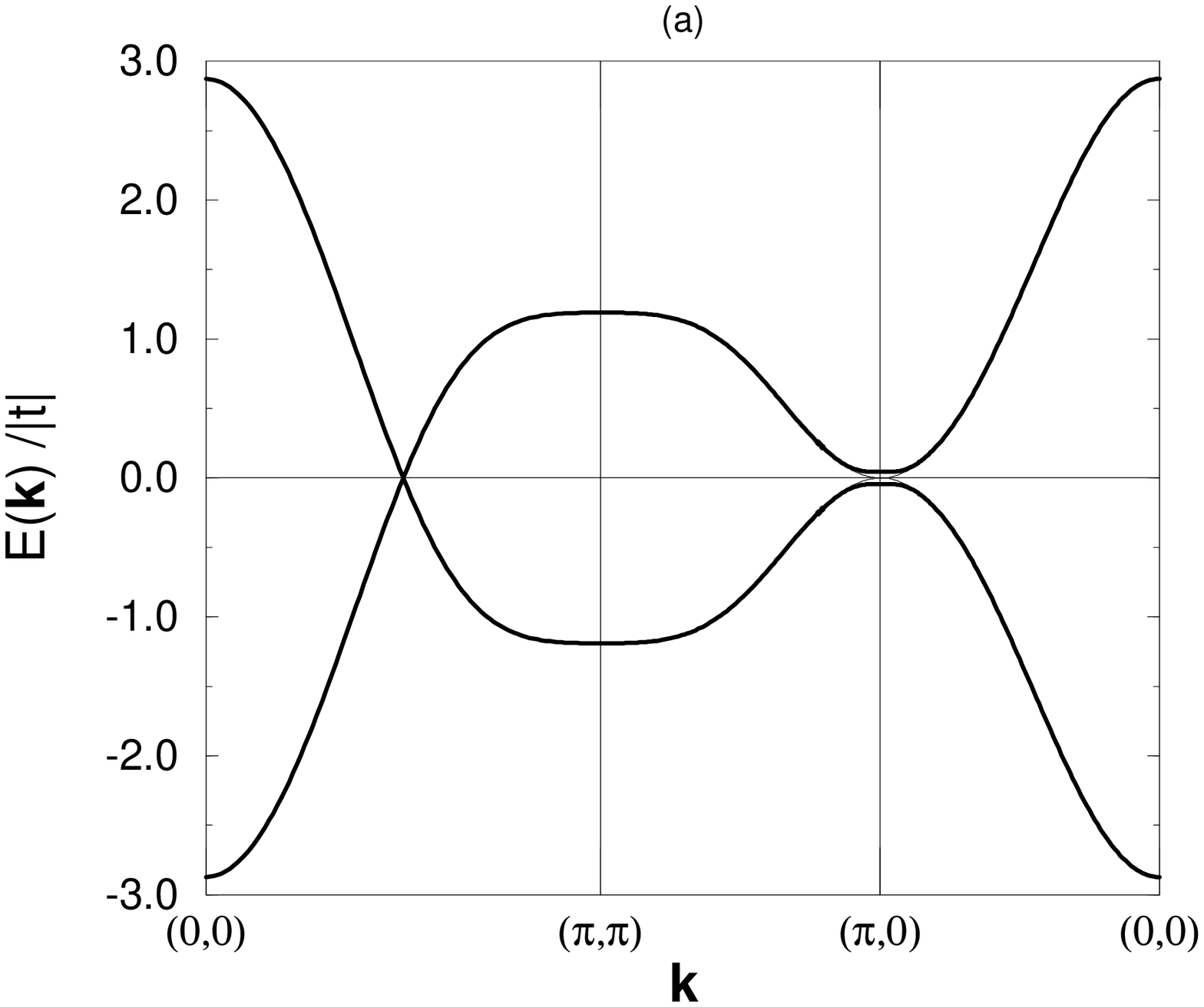}
\end{center}
\begin{center}
\leavevmode
\epsfxsize=4.2in
\epsffile{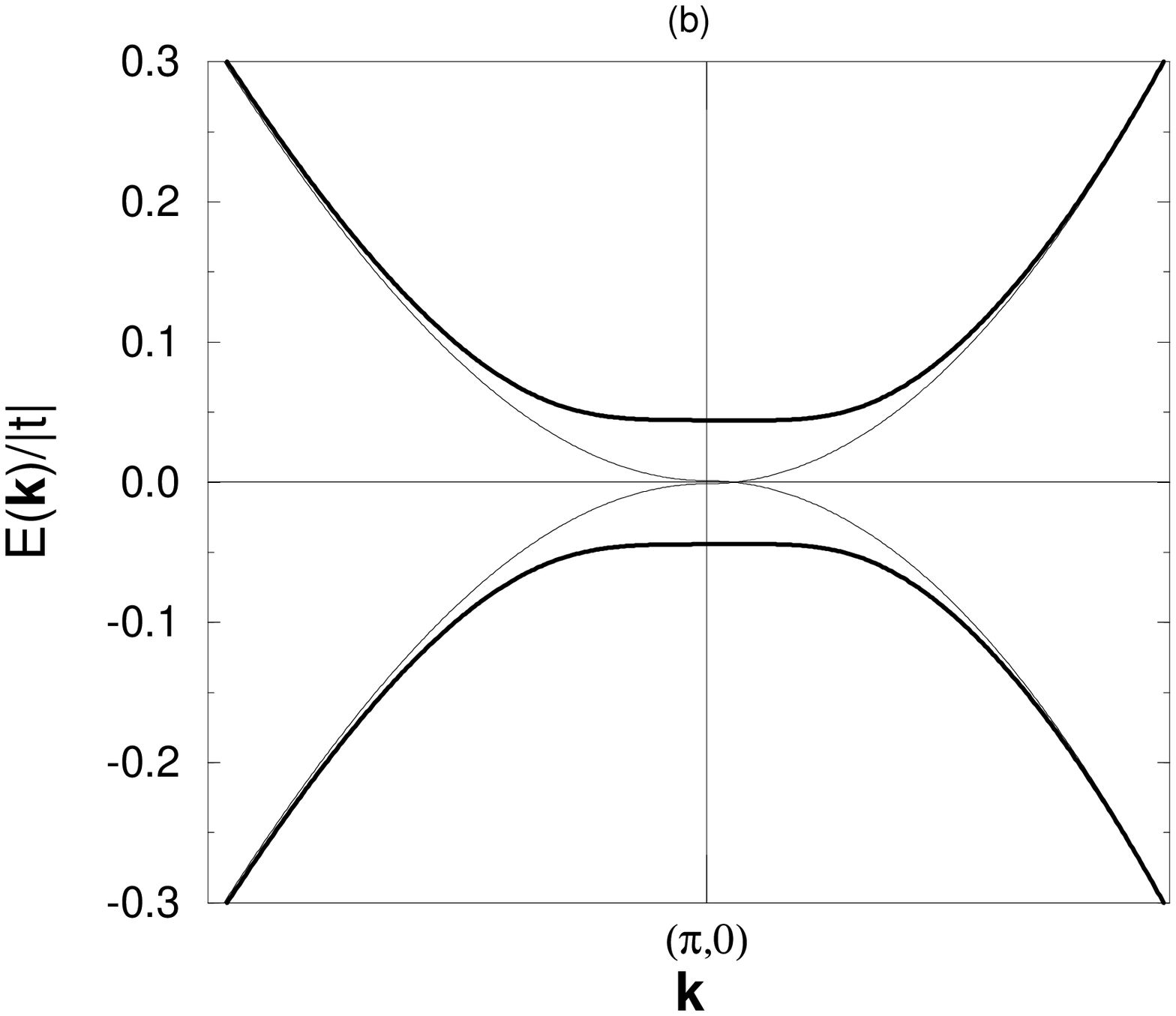}
\end{center}

\caption {(a) The 2 bands close to the Fermi surface in 
the superconducting state, calculated
using the factorization procedure.
 Note the gap in the neighbourhood
of the ($\pi,0$) point and the zero gap on the $k_x=k_y$ 
diagonal, reflecting the $d$-wave symmetry. Here $U=8|t|$ and
$<N>=0.795$ (nearly optimum doping).
(b) shows the neighbourhood of the $(\pi,0)$-point in detail.
The normal state electron and hole bands are shown in thin lines.} 
\label{figure:bandgap}
\end{figure}

The Green's function matrix must be calculated from 
Eq. (\ref{eq:G}). Here
we concentrate on the element $G_{13} = \langle \langle 
c_{{\vec k} \sigma}
; c_{-{\vec k} -\sigma} \rangle \rangle$ from which we can 
determine the
pair wavefunction $\langle c_{i -\sigma} c_{l \sigma} \rangle$. 
The poles of
the Green's function give the quasiparticle bands. It is convenient 
to
partition the matrix $\left(\omega\bf{N}_4 - \bf{E}_4\right)$ into 
four 2x2
matrices: \begin{equation}
\omega\bf{N}_4 - \bf{E}_4=\left[\begin{array}{cc}A&B\\
C&D\end{array}\right].
\end{equation}
Then
\widetext
\[
\bf{G}(\omega)=\bf{N}_4\left(\omega\bf{N}_4-\bf{E}_4\right)^{-1}
\bf{N}_4=
\]
\begin{equation}
\left[\begin{array}{cc}\bf{N}_2&\emptyset\\
\emptyset&\bf{N}_2\end{array}\right]
\left[\begin{array}{cc}
\left(A-BD^{-1}C\right)^{-1}&-A^{-1}B\left(D-CA^{-1}B\right)^{-1}\\
-D^{-1}C\left(A-BD^{-1}C\right)^{-1}&\left(D-CA^{-1}B\right)^{-1}
\end{array}\right]
\left[\begin{array}{cc}\bf{N}_2&\emptyset\\
\emptyset&\bf{N}_2\end{array}\right]
\end{equation}
\narrowtext and the calculation of $G_{13}$ only involves the upper
right-hand block $-A^{-1}B\left(D-CA^{-1}B\right)^{-1}$. It is now
straightforward to show that 
\begin{equation}
G_{13}(\omega) = - {\gamma_{\vec k} ( E_{12} - n E_{11})^2}
/{ D(\omega)} ,
\end{equation}
where
\begin{eqnarray}
D(\omega)&=&det\left(\bf{N}_4\omega - \bf{E}_4\right) \nonumber\\
&=&\left((\omega-E_{11})(\omega
n-E_{22})-(\omega n-E_{12})^2\right)\nonumber\\
&&\times\left((\omega+E_{11})(\omega
n+E_{22})-(\omega n+E_{12})^2\right)\nonumber\\
&&-\gamma_{\vec{k}}^*\gamma_{\vec{k}}(\omega^2-E_{11}^2).
\label{eq:Det}
\end{eqnarray}
$E_{11},E_{12}$ and $E_{22}$ may be substituted from 
Eq. (\ref{eq:E2}) and
we find 
\begin{equation}
G_{13}(\omega) = - \gamma_{\vec k} {( n(1-n)U)^2}/{D(\omega)}.
\label{eq:G13} 
\end{equation}
The quasiparticle bands in the superconducting state satisfy 
$D(\omega)=0$ and we denote them by $\omega=\pm E_{1\vec k}, 
\omega=\pm E_{2\vec k}$, with $ E_{2\vec k}>E_{1\vec k}>0$. Thus 
\begin{equation}
D(\omega) = n^2(1-n)^2 (\omega^2 - E_{1 {\vec k}}^2) 
(\omega^2 - E_{2{\vec k}}^2).\label{eq:D}
\end{equation}
When ${|\gamma_{\vec k}|}/{n(1-n)}\ll\sqrt{\xi_{2\vec{k}}^2
-\xi_{1\vec{k}}^2}$, we have 
\begin{equation}
E_{1\vec k}^2=\xi_{1\vec k}^2 +\frac{|\gamma_{\vec k}|^2}
{n^2 (1-n)^2}\frac{E_{11}^2-\xi_{1\vec k}^2}
{\xi_{2\vec k}^2-\xi_{1\vec
k}^2},\label{eq:E1k} 
\end{equation}
\begin{equation}
E_{2\vec k}^2=\xi_{2\vec k}^2 +\frac{|\gamma_{\vec k}|^2}
{n^2 (1-n)^2}\frac{\xi_{2\vec k}^2- E_{11}^2}
{\xi_{2\vec k}^2-\xi_{1\vec
k}^2},\label{eq:E2k} \end{equation}
and, from Eq. (\ref{eq:E2}), $E_{11}=\epsilon_{\vec k}+Un-\mu$.
Here $\xi_{1\vec k}$, $\xi_{2\vec k}$ with  $\xi_{1\vec k}<
\xi_{2\vec k}$ are the quasiparticle 
bands in the normal state given by Eq. (\ref{eq:xi12}). 
The form of $E_{1\vec k}^2$ is familiar from BCS-theory and the 
superconducting gap at points $\vec{k}$ 
on the normal state Fermi surface $\xi_{1 \vec k}=0$ is given by 
$|\Delta_{\vec k}|$
where, combining Eq.~(\ref{eq:gammak}) and Eq. (\ref{eq:E1k}), 
\begin{equation}
\Delta_{\vec k}= \frac{g\left(\cos k_xa-\cos k_ya\right)}{n(1-n)}.
\frac{(\epsilon_{\vec
k}+Un-\mu)}{\xi_{2\vec k}}.\label{eq:Deltak} \end{equation}
The $\vec{k}$-dependence of $\Delta_{\vec k}$ is entirely due to 
the $\left(\cos k_xa-\cos
k_ya\right)$-factor, since the second factor is constant over 
the Fermi surface.  So, over the
Fermi surface, 
\begin{equation}
\Delta_{\vec{k}}=G\left(\cos k_{x}a - cos k_{y}a\right),
\end{equation}
with
\begin{equation}
G=\frac{g(\epsilon_{\vec{k_F}}+Un-\mu)}{n(1-n)\xi_{2\vec{k_F}}}.
\label{eq:gap amplitude}
\end{equation}
We call $G$ the gap amplitude. Recall that we call $g$, 
introduced in 
Eq. (\ref{eq:g}), the gap-function amplitude.
$G$ tends to $g/(1-n)$ as $U\rightarrow\infty$. Examples 
of quasiparticle bands in 
the superconducting state are given later in 
Fig.~\ref{figure:bandgap}, 
with the gap-function amplitude $g$ determined by a 
method described in
 Sec.~\ref{section:gap}.

In calculations which relate to the superconducting
state it is reasonable to assume that $W_{\vec k}$, 
which appears in the matrix
$\bf{E}_2$ [whose elements appear in Eq. (\ref{eq:G13})], 
takes the values it
has in the normal state (where $\gamma_{\vec k} = 0$) at $T=0$. 
Thus the effect
of superconductivity on the quasiparticle bands, and the 
temperature dependence
of these bands, is entirely due to the gap function 
$\gamma_{\vec{k}}$. The
correlation function $\left<c_{-\vec{k}-\sigma}
c_{\vec{k}\sigma}\right>$ is
related to $G_{13}(\omega)$ by the use of Eq. (\ref{eq:Fw}) 
so that, using Eq. (\ref{eq:G13}) and Eq.~(\ref{eq:D}), 
\begin{equation}
\left<c_{-\vec{k}-\sigma}c_{\vec{k}\sigma}\right>=\frac{1}{2\pi i}
\oint f(\omega) \frac{-\gamma_{\vec{k}}U^2}
{(\omega^2-E_{1\vec{k}}^2)(\omega^2-E_{2\vec{k}}^2)}d\omega.
\end{equation}
Hence 
\begin{equation}
\left<c_{-\vec{k}-\sigma}c_{\vec{k}\sigma}\right>=
-\gamma_{\vec{k}}U^2 
F\left(E_{1 {\vec k}},E_{2 {\vec k}}\right),\label{eq:c-kckF}
\end{equation}
where
\begin{equation}
F(a,b)=\frac{1}{2\left(b^2-a^2\right)}\left[\frac{\tanh 
\left(\frac{1}{2}{\beta a}\right)}{a}-\frac{\tanh 
\left(\frac{1}{2}{\beta b}\right)}{b}\right].
\end{equation}
At $T=0$ $(\beta\rightarrow\infty)$ this becomes
\begin{equation}
\left<c_{-\vec{k}-\sigma}c_{\vec{k}\sigma}\right>=\frac
{-\gamma_{\vec{k}}U^2}{2E_{1\vec k}E_{2\vec k}
\left(E_{1\vec k}+E_{2\vec k}
\right) }.
\end{equation}
The pair wave function in real space may then be calculated 
using the relation
\begin{equation}
\left< c_{i-\sigma} c_{l\sigma}\right>=\frac{1}{L}\sum_{\vec k}
e^{i\vec{k}.\left(\vec{R}_i-\vec{R}_l\right)}
\left<c_{-\vec{k}-\sigma}c_{\vec{k}\sigma}\right>.\label{eq:cicl}
\end{equation}

\section{The gap equation}\label{section:gap}

In the $d$-wave case the gap is determined by the gap-function 
$\gamma_{\vec k}$
and we need a self-consistent way of finding the gap-function 
amplitude $g$ which appears
in Eq. (\ref{eq:gammak}). As pointed out in the previous section 
the gap
is not determined by a simple pair wave 
function $\left< c_{i-\sigma} c_{l\sigma}\right>$, 
but by the
correlation function $\gamma_{il}$ given by 
Eq. (\ref{eq:gammail}), which
may be written as
\begin{equation} \gamma_{il} =
\left<\left( c_{i-\sigma}^\dagger c_{l-\sigma}+ 
c_{l\sigma}^\dagger c_{i\sigma}\right) c_{i-\sigma} 
c_{l\sigma}\right>.\label{eq:gamma}
\end{equation}
In the $d$-wave case this may also be written as 
\begin{equation}
\gamma_{il}=
\left< c_{i\sigma}^\dagger c_{l\sigma} c_{l-\sigma} 
c_{i\sigma}\right> + \left< c_{l\sigma}^\dagger c_{i\sigma} 
c_{i-\sigma} c_{l\sigma}\right>,\label{eq:gammads}
\end{equation}
which shows explicitly that $\gamma_{il}=\gamma_{li}$.
This is not expressible directly in terms of our Green's 
functions and various
possibilities suggest themselves. A satisfactory procedure 
might be to extend
the formalism to two-particle Green's functions as suggested 
in connection with
the correlation functions appearing in $w_1$, Eq. (\ref{eq:w1}). 
This remains to be
investigated.  Alternatively we
can follow the procedure we adopted, following Roth, for the latter
correlation functions and introduce additional Green's 
functions containing
one- and three-particle operators. This is the method 
we shall use here, since it is 
consistent with our treatment of the normal state which, 
by comparison with Monte
Carlo results, has been shown to work well. However there 
is one complication.
Whereas for the correlation functions appearing in the 
normal paramagnetic state the
procedure is unambiguous,\cite{roth} this is not so for 
$\gamma_{il}$.
The four-operator correlation function may be related to a 
Green's function
containing a one- and three-particle operator in four ways and 
in the case of
 $\gamma_{il}$ these are distinct and yield different results.
It is  sufficient to consider one of the correlation functions 
in Eq. (\ref{eq:gammads}).
Thus $\left< c_{l\sigma}^\dagger c_{i\sigma} c_{i-\sigma} 
c_{l\sigma}\right>$ may be calculated using any of the following 
Green's functions:\\
$\;\;\;\;\;$(a) $\left<\left< c_{i\sigma}; c_{l\sigma}^\dagger 
c_{i-\sigma} c_{l\sigma}\right>\right>,\;\;\;\;\;\;\;\;\;\;$ 
(b)$\left<\left< c_{i-\sigma}; c_{l\sigma}^\dagger c_{i\sigma} 
c_{l\sigma}\right>\right>$,\\
$\;\;\;\;\;$(c) $\left<\left< c_{l\sigma}; c_{l\sigma}^\dagger 
c_{i\sigma} c_{i-\sigma}\right>\right>,\;\;\;\;\;\;\;\;\;\;$ 
(d)$\left<\left< c_{l\sigma}^\dagger; c_{i\sigma} c_{i-\sigma} 
c_{l\sigma}\right>\right>$.\\
Use of (a) or (b) splits up the product $ c_{i\sigma} 
c_{i-\sigma}$ and therefore does not preserve the 
important result that $\gamma_{il}\rightarrow 0$ when 
site double occupation is 
suppressed as $U\rightarrow\infty$. Consequently both of 
these procedures,
which yield similar but not identical results, fail at 
large $U$ and are generally
expected to overestimate the gap. This result is closely 
related to the 
factorization $\left< c_{l\sigma}^\dagger c_{i\sigma} 
c_{i-\sigma} c_{l\sigma}\right>=\left< c_{l\sigma}^\dagger 
c_{i\sigma}\right>\left< c_{i-\sigma} c_{l\sigma}\right>$
 and in fact use of (a) or (b) is exactly equivalent to this
factorization as $U\rightarrow\infty$. 
Rather than use (a) or (b) precisely we develop the 
related factorization method in 
Sec.~\ref{subsection:factor}. This has the advantage of 
simplicity and shows clearly 
the structure of the gap equation and why $d$-wave pairing 
occurs and $p$-wave pairing does not. 
Although it fails for large $U$, we believe it is useful in 
providing an upper estimate
of the gap and $T_c$ at finite, intermediate $U$. 

Use of the Green's functions (c) or (d) preserves the property
$\gamma_{il}\rightarrow 0$ as $U\rightarrow\infty$ and is 
therefore the prefered procedure  for large
$U$. At smaller $U$ it seems likely that the correct results 
may lie between those obtained from
(c) or (d) and those from (a), (b), or the factorization procedure.
We therefore regard the gap and $T_c$ obtained using 
(c) or (d) as a lower estimate of those quantities.
It is remarkable that although analyses using (c) and (d) 
are formally different the numerical results for 
the gap are almost identical. In 
Sec.~\ref{subsection:largeUdecoupling} we will therefore 
give details of the gap equation for case (c) only. 
Of the four possibilities (c)
has the advantage of giving the  correct $U\rightarrow\infty$ 
limit for the correlation function
and of a more balanced distribution of creation and 
destruction operators than (d).

There is a reassuring consistency in that all the procedures 
discussed yield $d$-wave pairing,
not $p$-wave pairing, and in all cases the gap and $T_c$ as 
function of site occupation $\left<N\right>$ 
are sharply peaked at an optimum doping corresponding to the 
square Fermi surface $\epsilon_{\vec k}=0$.
These results are presented in Secs.~\ref{subsection:factor} 
and~\ref{subsection:largeUdecoupling}
for the two basic methods which provide upper and lower 
estimates of the magnitude of the gap
and $T_c$.

\subsection{The factorization procedure}\label{subsection:factor}
The factorization procedure approximates $\gamma_{il}$, 
given by Eq. (\ref{eq:gamma}), as
\begin{equation}
\gamma_{il}=
\left(\left< c_{i-\sigma}^\dagger c_{l-\sigma}\right>
+\left< c_{l\sigma}^\dagger c_{i\sigma}\right>\right) 
\left< c_{i-\sigma} c_{l\sigma}\right>.\label{eq:gapfac}
\end{equation}
The symmetry $\gamma_{il}=\gamma_{li}$ is retained but, 
as discussed above, the products
$ c_{i\sigma} c_{i-\sigma}$ and $ c_{l-\sigma} c_{l\sigma}$ 
are split up so that in general $\gamma_{il}$ does not tend 
to zero when
site double occupation is suppressed as $U\rightarrow \infty$. 
However the averages $\left< c_{l-\sigma} c_{l\sigma}\right>$
are always zero due to $d$-wave symmetry, so that double 
occupation within a pair is suppressed,
but not between pairs. In the special case of half-filling 
($\left<N\right>=1$), 
$\gamma_{il}\rightarrow 0$ as $U\rightarrow \infty$, 
because the correlations
$\left< c_{i-\sigma}^\dagger c_{l-\sigma}\right>$ and 
$\left< c_{l\sigma}^\dagger c_{i\sigma}\right>$ tend to zero.
We may rewrite Eq. (\ref{eq:gapfac}) as
\begin{equation}
\gamma_{il}=2n_1\left< c_{i-\sigma} c_{l\sigma}\right>,
\label{eq:gapfacn1}
\end{equation}
where
\begin{equation}
n_1=\left< c_{i-\sigma}^\dagger c_{l-\sigma}\right>=
\left< c_{l\sigma}^\dagger c_{i\sigma}\right>.\label{eq:n1}
\end{equation}
It is straightforward to show that 
\begin{equation}
n_1t=\frac{1}{zL}\sum_{\vec k}\epsilon_{\vec k}n_{\vec k},
\label{eq:n1t}
\end{equation}
where $n_{\vec k}=\left<c_{\vec{k}\sigma}^\dagger 
c_{\vec{k}\sigma}\right>$ and
$z$ is the number of nearest neighbours, i.e. $z=4$ 
for the 2D square lattice.

We are now in a position to find the equation which determines 
the gap-function
amplitude $g$ in Eq. (\ref{eq:gammak}) self-consistently.
From Eqs. (\ref{eq:gammak}) and (\ref{eq:gapfacn1}) we get 
\begin{equation}
\gamma_{\vec k}=2n_1 t \sum_{\left<l\right>_i} \cos
\left[\vec{k}.\left(\vec{R}_i-\vec{R}_l\right)\right]
\left< c_{i-\sigma} c_{l\sigma}\right>. 
\end{equation}
By using Eqs. (\ref{eq:c-kckF}) and (\ref{eq:cicl}) we find:
\widetext
\begin{equation}
\gamma_{\vec k}=-\frac{2n_1 tU^2}{L} \sum_{\left<l\right>_i} 
\cos \left[\vec{k}.\left(\vec{R}_i-\vec{R}_l\right)\right]
\sum_{\vec q} \gamma_{\vec{q}}
\cos \left[\vec{q}.\left(\vec{R}_i-\vec{R}_l\right)\right]
F\left(E_{1\vec{q}},E_{2\vec{q}}\right).\label{eq:gapeqder1}
\end{equation}
\narrowtext
Clearly the quasiparticle energies $E_{1\vec{q}},\; E_{2\vec{q}}$ 
are symmetric in $q_x$ and $q_y$, because they satisfy the equation 
$D(\omega)$=0, where $D(\omega)$ is defined by Eq. (\ref{eq:Det}) 
and $|\gamma_{\vec q}|^2$ and the elements of $\bf{E}_2$ have 
this symmetry.
Now substituting Eq. (\ref{eq:gammak}), 
\begin{equation}
\gamma_{\vec q}= g \; \left(\cos q_xa -\cos q_ya\right)\;,
\label{eq:g*cos}
\end{equation}
into Eq. (\ref{eq:gapeqder1}) and exploiting the 
symmetry between $q_x$ and $q_y$, we
find: 
\begin{equation}
\gamma_{\vec k}=- \gamma_{\vec k} \frac{2n_1 tU^2}{L}
\sum_{\vec{q}} \left(\cos q_xa - \cos q_ya\right)^2 
F\left(E_{1\vec{q}},E_{2\vec{q}}\right).
\end{equation} 
Thus for a non-zero solution $\gamma_{\vec k}$ must 
satisfy the gap equation
\begin{equation}
1=-\frac{2n_1 tU^2}{L}\sum_{\vec{k}} 
\left(\cos k_xa - \cos k_ya\right)^2 
F\left(E_{1\vec{k}},E_{2\vec{k}}\right), \label{eq:gapequation}
\end{equation}
where $\gamma_{\vec k}$ is contained in the quasiparticle energies
$E_{1\vec{k}},E_{2\vec{k}}$. At $T=0$, $F(a,b)$ takes the 
form $[2ab(a+b)]^{-1}$, so that
the gap equation becomes: 
\begin{equation} 
1=-\frac{n_1tU^2}{L}\sum_{\vec k}\frac{\left(\cos k_xa-\cos
k_ya\right)^2} {E_{1\vec{k}}E_{2\vec{k}}\left(E_{1\vec{k}}
+E_{2\vec{k}}\right)}.
\label{eq:gapequationT=0} 
\end{equation} 
Since $\sum_{\vec k}\epsilon_{\vec k}=0$ and $n_{\vec k}$ is a
monotonically decreasing function of $\epsilon_{\vec k}$, 
it follows from Eq. (\ref{eq:n1t})
that $n_1t<0$. Also $E_{2\vec{k}}>E_{1\vec{k}}>0$, so that 
a solution with $g\ne 0$ may occur.
At finite temperature $T$ Eq. (\ref{eq:gapequation}) 
determines $g$ as a function of temperature,
and hence the critical temperature $T_c$ where $g=0$. 
Eq. (\ref{eq:gapequationT=0}) has a
rather BCS-like form, particularly if we allow ourselves 
to consider large $U$ where $E_{2\vec
k}\sim U$ and the  factor $E_{2\vec k}\left(E_{1 \vec k}
+E_{2 \vec k}\right)$ in the
denominator in Eq. (\ref{eq:gapequationT=0}) cancels with 
the factor $U^2$ in the  numerator.
In this approximation the superconductivity is driven by the 
kinetic energy term $-n_1t$, but
for large $U$, where the  approximation of 
Sec.~\ref{subsection:largeUdecoupling} is more
appropriate, this is effectively replaced by a 
quantity of order ${t^2}/{U}$. 

In the case of $p$-wave pairing, which was briefly
considered in Sec.~\ref{section:scstate}, the equation 
corresponding to
Eq. (\ref{eq:gapequationT=0}) is of the form 
\begin{equation} 1=\frac{n_1tU^2}{L}\sum_{\vec
k}\frac{\left(\sin k_xa  \pm \sin k_ya\right)^2}
{E_{1\vec{k}}E_{2\vec{k}}\left(E_{1\vec{k}}+E_{2\vec{k}}\right)} .
\label{eq:pgapequationT=0}
\end{equation} The right-hand side is now negative and 
no solution is possible. The change of sign from
the $d$-wave case is due to the factor $i$ in Eq. (\ref{eq:E24p}), 
which does not appear
in Eq. (\ref{eq:E24ds}). Thus $p$-wave superconductivity 
does not occur. 

\begin{figure}

\begin{center}
\leavevmode
\epsfxsize=6.5in
\epsffile{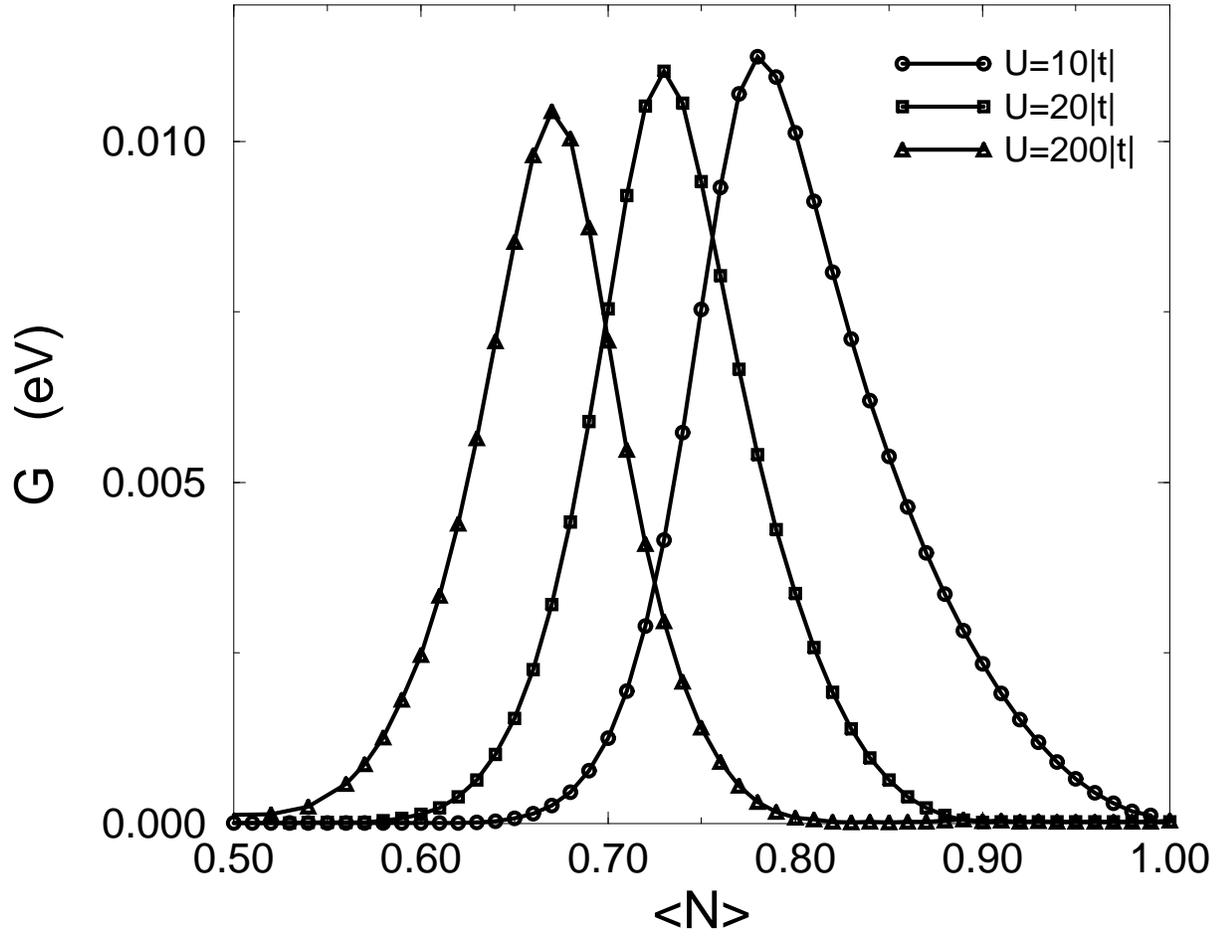}
\end{center}

\caption{The gap amplitude $G$ as a function of the 
occupation $<N>$ for several ratio's $U/|t|$, 
calculated using the factorization procedure. ($|t|=0.5 eV$).}
\label{figure:gap amplitude}
\end{figure}

As mentioned earlier, in calculating superconducting
properties such as the gap amplitude and its temperature  
dependence we neglect the
unimportant temperature dependence of $n_1$ and $W_{\vec k}$ 
and evaluate  these with $g=0$
and $T=0$. The $d$-wave gap amplitude $G$ at $T=0$ over the 
Fermi surface calculated from
Eqs. (\ref{eq:gapequationT=0}) and (\ref{eq:gap amplitude})  
is shown in
Fig.~\ref{figure:gap amplitude}  as a function of occupation 
$\left<N\right>$ for different
values of ${U}/{|t|}$. Clearly, from Eq. (\ref{eq:Deltak}) 
the maximum gap occurs at
points on the Fermi-surface where $k_x=0$ or $k_y=0$. For 
each ${U}/{|t|}$ the gap
maximum, corresponding to optimum doping, occurs for the 
occupation which has the square Fermi
surface $\epsilon_{\vec k}=0$. The flat saddle-point feature 
of the normal state
quasiparticle bands (Fig.~\ref{figure:bands}) around $(\pi,0)$ 
then lies precisely at the
Fermi level.  If the occupation is increased 
(underdoping with holes) the saddle-point lies
below the Fermi level and the Fermi surface is 
a closed hole-like one around $(\pi,\pi)$. For
overdoping the saddle-point lies above the Fermi 
level and the  Fermi surface is a closed
electron-like one around $(0,0)$. This situation is 
unchanged in the approximation of
Sec.~\ref{subsection:largeUdecoupling}, which is more 
appropriate for large $U$. It has
close similarities with the van Hove scenario of 
Newns {\it et al}.~\cite{newns} and has
implications for transport properties which we explore 
in the discussion section.

Quasiparticle bands in the superconducting state are 
obtained by solving the equation
$D(\omega)=0$, where $D(\omega)$ is given by Eq. (\ref{eq:Det}) 
and the bands are given to a
good approximation by Eqs. (\ref{eq:E1k}) and (\ref{eq:E2k}). 
Examples of these bands are
shown in Fig.~\ref{figure:bandgap}. The upper Hubbard bands 
$\omega = \pm E_{2\vec k}$ are too
far from the Fermi level to appear in this figure. 

\begin{figure}
\begin{center}
\leavevmode
\epsfxsize=5in
\epsffile{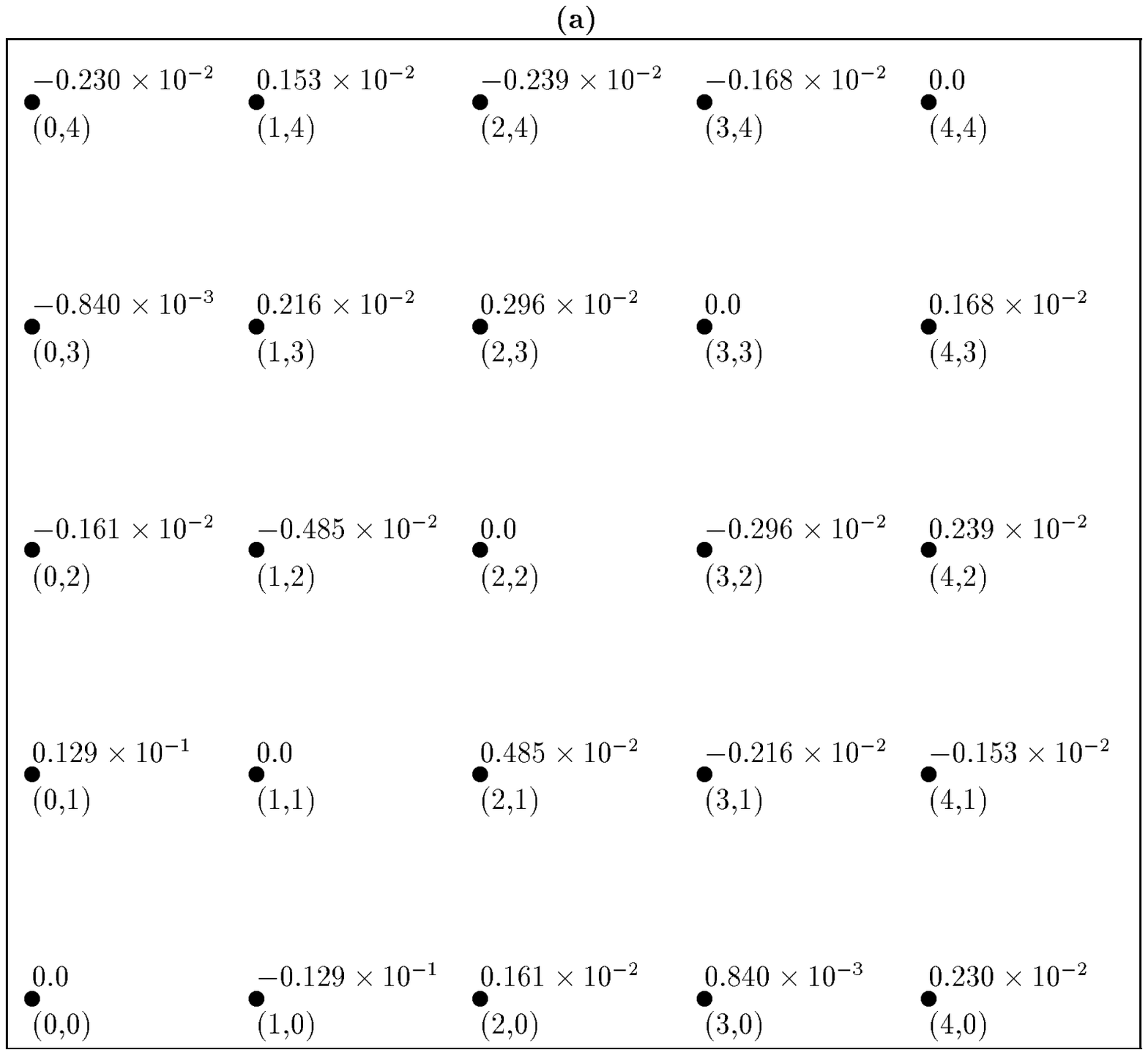}
\end{center}
\begin{center}
\leavevmode
\epsfxsize=5in
\epsffile{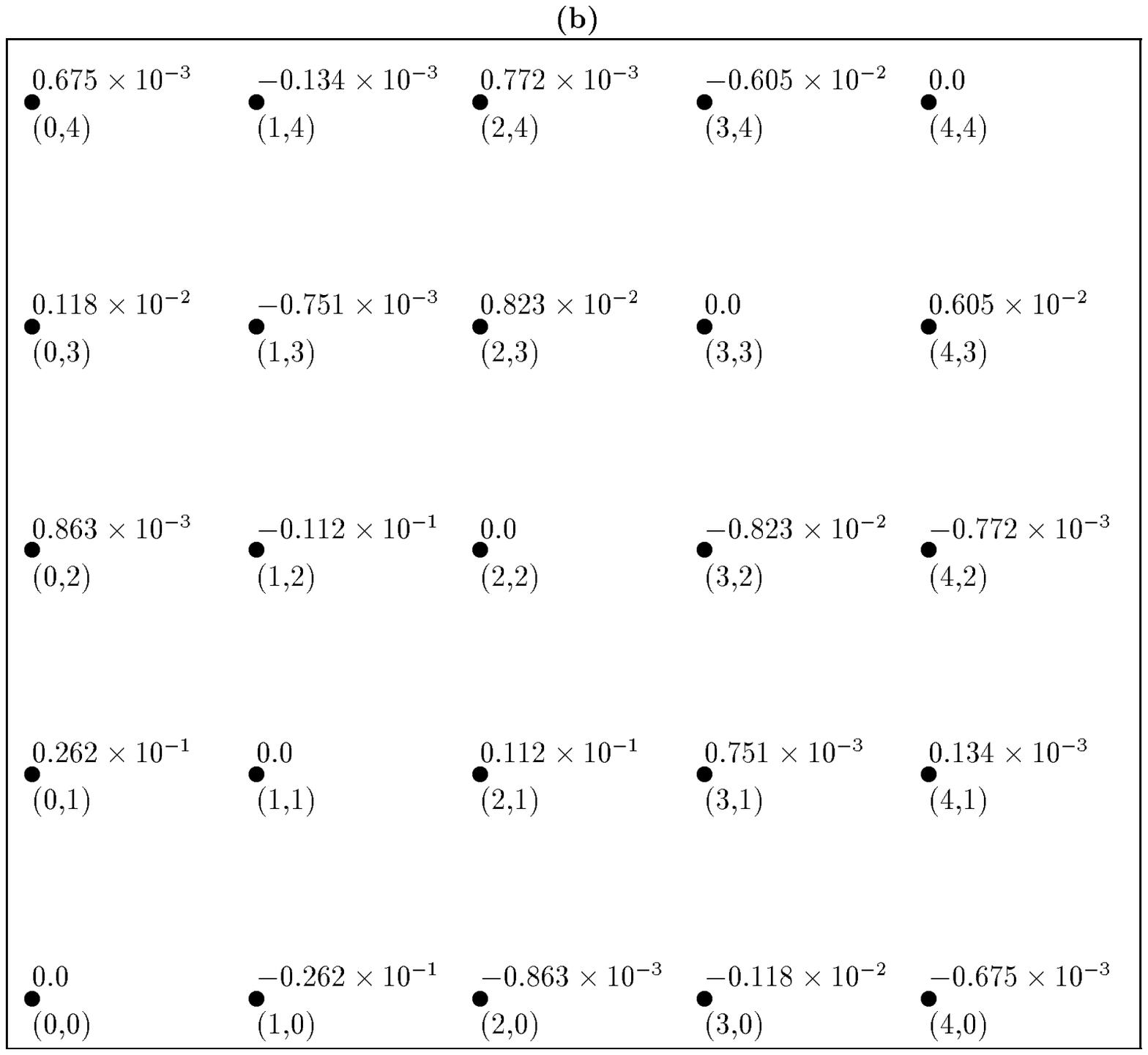}
\end{center}
\begin{center}
\leavevmode
\epsfxsize=5in
\epsffile{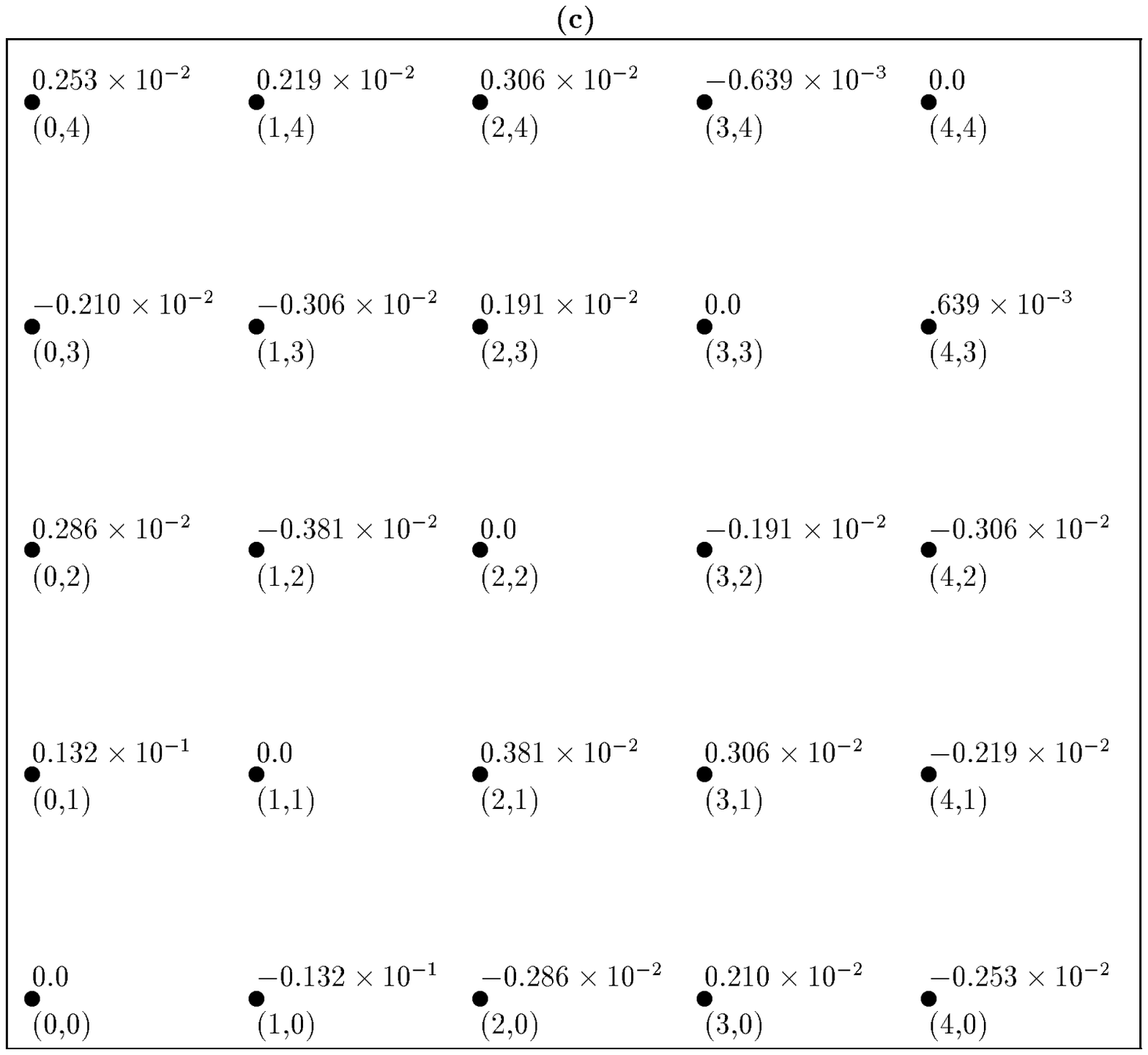}
\end{center}

\caption{Values of the pair wave function 
$\left<c_{i -\sigma}c_{l\sigma}\right>$ for several lattice vectors
$\vec{R}_i - \vec{R}_l$. The pair wave function value is printed
above the lattice points and the lattice vector is given, as a
 coordinate pair in units of the lattice constant, below the 
lattice points. $U=12|t|$. (a) $<N>=0.72$ (overdoping), 
(b) $<N>=0.76$ (optimum doping), 
and (c) $<N>=0.82$ (underdoping).}
\label{figure:pair}
\end{figure}

In Fig.~\ref{figure:pair} we print the values of the pair wave 
function $\left<c_{i -\sigma}c_{l\sigma}\right>$, calculated
from Eq.~(\ref{eq:cicl}), for several $\vec{R}_i -\vec{R}_l$ on the 
lattice. It turns out that for optimum doping the pair wave
function is strongly peaked if $i$ and $l$ are neighbouring sites.
Away from optimum doping this peak is less pronounced and the 
pair wave function extends over several lattice sites. 
This behaviour of the pair wave function reflects the local nature
of the pairing mechanism.

\begin{figure}

\begin{center}
\leavevmode
\epsfxsize=6.5in
\epsffile{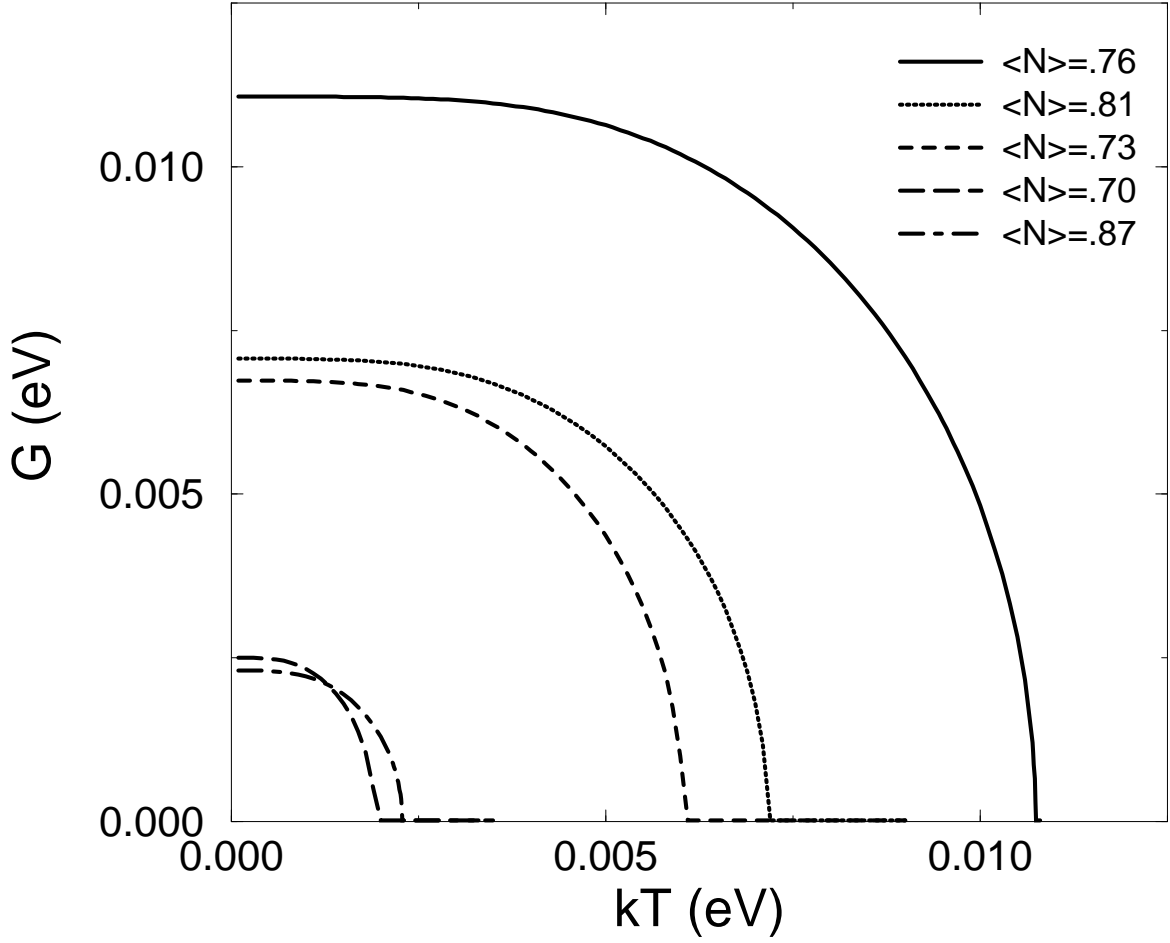}
\end{center}

\caption{The gap amplitude $G$ as a function of temperature 
for $U=12|t|=6eV$ for several
occupations $<N>$, calculated using the factorization procedure.
Note that the maximum gap $\Delta=4G$ at optimum 
doping $\delta_c$ (=0.24 here). }
\label{figure:GT}
\end{figure}

The temperature dependence of the
gap-function amplitude $g(T)$ is calculated from 
Eq. (\ref{eq:gapequation}) and in
Fig.~\ref{figure:GT} results for $G(T)$ are shown 
for ${U}/{|t|}=12$ and various
occupations including the optimum doping case of 
$\left<N\right>=0.76$. It is clear from
Fig.~\ref{figure:GT} that $2\Delta_{max}/k_B T_c \sim 4$ 
for optimum doping and
under-doping, but larger for overdoping. The units of $g$ 
and $k_B T$ in
Figs.~\ref{figure:gap amplitude} and~\ref{figure:GT} are $2|t|$, 
so that if $|t|$ takes a
reasonable value of $0.5$ eV, the units are eV. Then $T_c$ 
at optimum doping for
${U}/{|t|}=12$ is $125$ K. As discussed earlier we regard 
these results for the gap
and $T_c$ as upper bound estimates. They are strongly reduced 
in the approximation of the next
section, which we believe yields lower bound estimates, 
but the qualitative behaviour is
unchanged.

\subsection{A gap equation valid for 
large $U$}\label{subsection:largeUdecoupling}

\begin{figure}
\begin{center}
\leavevmode
\epsfxsize=4.2in
\epsffile{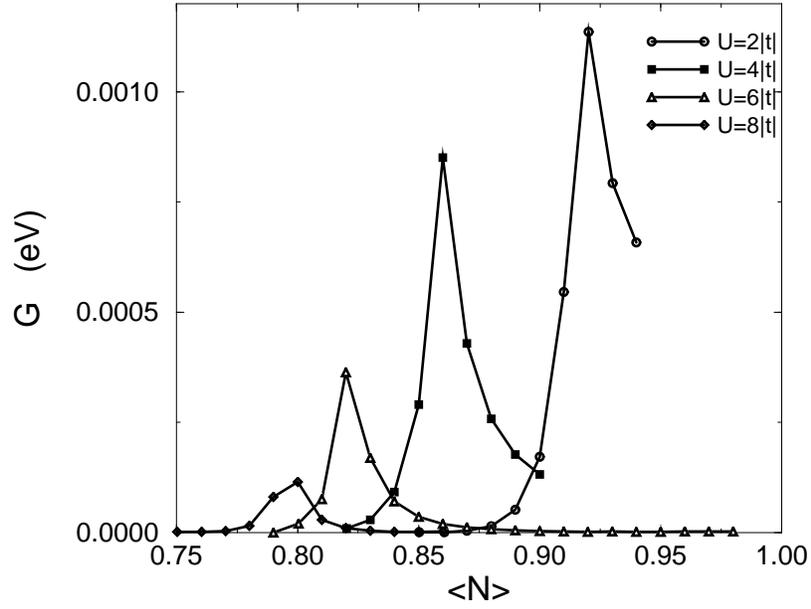}
\end{center}

\caption{The gap amplitude $G$ as a function of the 
occupation $<N>$ for several ratio's $U/|t|$, calculated 
using a decoupling of the gap-function
that is correct in the limit $U \rightarrow \infty$. 
($|t|=0.5 eV$).}
\label{figure:gap amplitudeII}
\end{figure}

\begin{figure}
\begin{center}
\leavevmode
\epsfxsize=4.2in
\epsffile{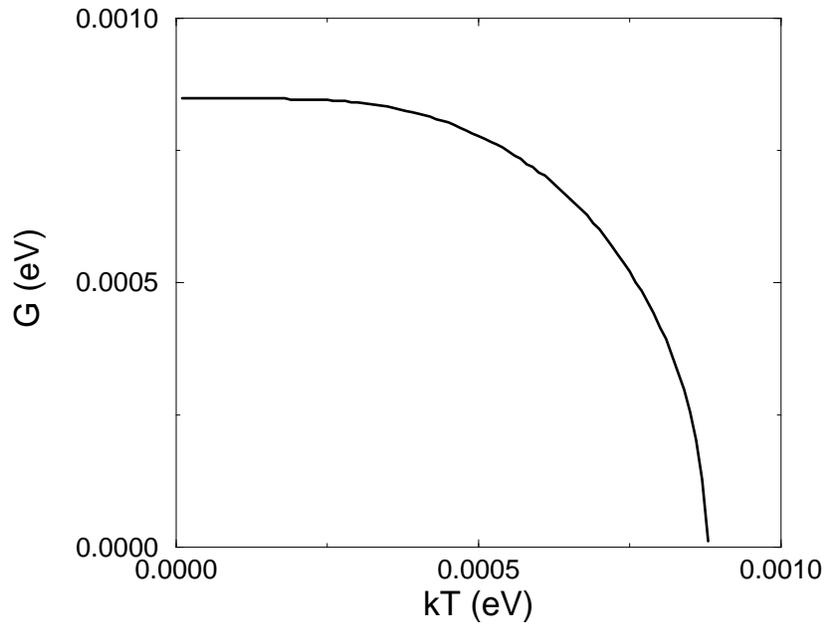}
\end{center}

\caption{The gap amplitude $G$ as a function of 
temperature for $U=4|t|=2eV$ for optimum doping $\delta_c=0.14$,
calculated using a decoupling of the gap function that 
is correct in the limit $U \rightarrow \infty$.} 
\label{figure:GTII}
\end{figure}

As discussed above, we obtain a gap equation by 
writing the correlation functions which appear in 
Eq. (\ref{eq:gammads}) as 
\begin{equation}
\left< c_{l\sigma}^\dagger c_{i\sigma} c_{i-\sigma} 
c_{l\sigma}\right>=\left<\hat{B}_{li} c_{l\sigma}\right>,
\end{equation}
with
\begin{equation}
\hat{B}_{li}= c_{l\sigma}^\dagger c_{i\sigma} 
c_{i-\sigma},
\end{equation}
and thus expressing it in terms of the Green's function 
$\left<\left< c_{l\sigma};\hat{B}_{li}\right>\right>$.
From Eq. (\ref{eq:gammads}) we have 
\begin{equation}
\gamma_{il}=\left<\hat{B}_{il} c_{i\sigma}\right>+
\left<\hat{B}_{li} c_{l\sigma}\right>.\label{eq:Bilci+Blicl}
\end{equation}
It is convenient to introduce Bloch operators by writing
\begin{equation}
\left<\hat{B}_{li} c_{l\sigma}\right>=L^{-1/2}
\sum_{\vec k}e^{-i\vec{k}.\vec{R}_l}
\left<\hat{B}_{li}c_{\vec{k}\sigma}\right>.\label{eq:Bck}
\end{equation}
Also, using Eqs. (\ref{eq:GtildeAB}) and (\ref{eq:Fw}) 
and noting that $\hat{A}_{1\vec{k}\sigma}=c_{\vec{k}\sigma}$
we have
\begin{equation}
\left<\hat{B}_{li}c_{\vec{k}\sigma}\right>=\sum_m I_m (\vec{k}) 
\left<\left[\hat{A}_{m\vec{k}\sigma},\hat{B}_{li}\right]_+\right>,
\label{eq:Blick}
\end{equation}
where
\begin{equation}
I_{m}(\vec k)=\frac{1}{2\pi i}\oint f(\omega)
\tilde{G}_{1m}(\vec{k},\omega)d\omega.\label{eq:Im}
\end{equation}
It is straightforward to show that
\begin{eqnarray}
\left<\left[\hat{A}_{1\vec{k}\sigma},\hat{B}_{li}\right]_
+\right>&=&0,\label{eq:A1kBli}\\
\left<\left[\hat{A}_{2\vec{k}\sigma},\hat{B}_{li}\right]_
+\right>&=&L^{-1/2}
e^{i\vec{k}.\vec{R}_l} 
\left< n_{l-\sigma} c_{i\sigma} c_{i-\sigma}\right>,
\label{eq:A2kBli}\\
\left<\left[\hat{A}_{3\vec{k}\sigma},\hat{B}_{li}\right]_+\right>&=&
L^{-1/2}e^{i\vec{k}.\vec{R}_i} n_1,\label{eq:A3kBli}\\
\left<\left[\hat{A}_{4\vec{k}\sigma},\hat{B}_{li}\right]_+\right>&=&
L^{-1/2}e^{i\vec{k}.\vec{R}_i}\left(n_1-m_1\right)
\nonumber\\
&&+ L^{-1/2}e^{i\vec{k}.\vec{R}_l}
\left< c_{l-\sigma}^\dagger c_{l\sigma}^\dagger c_{i\sigma} 
c_{i-\sigma}\right>
,\label{eq:A4kBli}
\end{eqnarray}
where $n_1$ is defined by Eq. (\ref{eq:n1}) and, similarly,
\begin{equation}
m_1=\left< c_{l\sigma}^\dagger n_{i-\sigma} 
c_{i\sigma}\right>=\left< c_{i\sigma}^\dagger 
n_{l-\sigma} c_{l\sigma}\right>.\label{eq:m1}
\end{equation}
So $m_1$ can be calculated straightforwardly from $G_{12}$ 
and it is sufficient to 
use its value in the normal
state at $T=0$. This may be written as
\begin{equation}
m_1t=\frac{n}{zL}\sum_{\vec k}\epsilon_{\vec k}
\frac{W_{\vec k}-\xi_{1\vec k}}{\xi_{2\vec k}
-\xi_{1\vec k}}f(\xi_{1\vec k})
\end{equation}
in the case when only the lower quasiparticle band is occupied. 
It is important to note that for large $U$, 
$m_1t\sim{t^2}/{U}$, whereas
$n_1t\sim t$. The expressions for $\tilde{G}_{1m}$ 
are obtained in a similar way to that used to find $G_{13}$ 
in Sec.~\ref{section:scstate}.
We find
\begin{eqnarray}
\tilde{G}_{12}&=&{Un^2(1-n)^2\left(\omega+\xi_{1 \vec k}\right)
\left(\omega+\xi_{2 \vec
k}\right)}/{D(\omega)},\label{eq:Gtilde12}\\ 
\tilde{G}_{13}&=&-\gamma_{\vec
k}{Un^2(1-n)\left(\omega+\epsilon_{\vec k}-\mu+U\right)}/
{D(\omega)},\label{eq:Gtilde13}\\
\tilde{G}_{14}&=&\gamma_{\vec k}{Un(1-n)\left(\omega
+\epsilon_{\vec k}-\mu+Un\right)}/{D(\omega)}.\label{eq:Gtilde14} 
\end{eqnarray} 
We note also that for $d$-wave symmetry
the correlation function on the right of Eq. (\ref{eq:A2kBli}) 
is equal to
$\left<\hat{B}_{li} c_{l\sigma}\right>$. It is also important 
note that, owing to the factors
$\gamma_{\vec k}$ in Eqs. (\ref{eq:Gtilde13}) and 
(\ref{eq:Gtilde14}), $I_3(\vec k)$ and
$I_4(\vec k)$ change sign under interchange of $k_x$ and $k_y$. 
Consequently on substituting
for the correlation functions on the right of Eq. (\ref{eq:Blick}) 
from
Eqs. (\ref{eq:A1kBli})-(\ref{eq:A4kBli}) and substituting  
the result in Eq. (\ref{eq:Bck}), we
find that the contribution from the second term on the right 
of Eq. (\ref{eq:A4kBli}) vanishes
on summing over $\vec k$. The result is 
\begin{equation}
\left<\hat{B}_{li} c_{l\sigma}\right>=\frac{C}{L}\sum_{\vec k}
e^{i\vec{k}.\left(\vec{R}_i -\vec{R}_l\right)}\left[I_3(\vec k) 
n_1 +I_4(\vec
k)(n_1-m_1)\right], 
\end{equation} 
where 
\begin{equation}
C=\left[1-L^{-1}\sum_{\vec q}I_2(\vec q)\right]^{-1}.
\end{equation}
It follows from Eqs. (\ref{eq:gammak}) and (\ref{eq:Bilci+Blicl}) 
that
\widetext 
\begin{equation}
\gamma_{\vec k}=\frac{2tC}{L}\sum_{\left<l\right>_i}
\cos\left[\vec{k}.(\vec{R}_i-\vec{R}_l)\right]\sum_{\vec q} 
\gamma_{\vec q}\cos\left[\vec{q}.(\vec{R}_i-\vec{R}_l)\right] 
\frac{I_3(\vec q) n_1 + I_4(\vec q) (n_1-m_1)}{\gamma_{\vec q}}.
\end{equation}\narrowtext
This is of the same form as Eq. (\ref{eq:gapeqder1}) and, 
on using Eq. (\ref{eq:g*cos}) as before, we find the gap equation
\begin{equation}
\frac{L}{2tC}=\sum_{\vec k} 
\left(\cos k_xa - \cos k_ya\right)^2\frac{I_3(\vec k)n_1
+ I_4(\vec k)(n_1-m_1)}{\gamma_{\vec k}} \label{eq:gapequationIII}
\end{equation}	
It is sufficient to evaluate the constant $C$ in the normal state 
at $T=0$ and we find
\begin{equation}
C=\left[1+UL^{-1}\sum_{\vec k}\left(\xi_{2\vec k}-\xi_{1\vec k}
\right)^{-1}\theta\left(-\xi_{1\vec k}\right)\right]^{-1},
\label{eq:gapequationT=0III}
\end{equation}
where $\theta(x)=1$ if $x>0$ and $=0$ otherwise.
As $U\rightarrow\infty$, $C\rightarrow 1-n$. At $T=0$, using
Eqs. (\ref{eq:Gtilde12})-(\ref{eq:Gtilde14}) 
and Eq. (\ref{eq:Im}), we obtain the gap equation
in the form \widetext
\begin{equation} 
1=-\frac{UtC}{n(1-n)L}\sum_{\vec k}\left(\cos k_xa - \cos k_ya
\right)^2\frac{\left[(1-n)n_1-m_1\right]
\left(\mu-\epsilon_{\vec k}\right)+Um_1n} 
{E_{1\vec k}E_{2\vec k}(E_{1\vec k}+E_{2\vec k})}.
\label{eq:gapequationT=0UinfIII} 
\end{equation}\narrowtext
This is similar in form to the gap equation, 
Eq. (\ref{eq:gapequationT=0}), 
obtained using the factorization method. Both terms in the 
numerator of the
last factor of Eq.~(\ref{eq:gapequationT=0UinfIII}) are of 
order $t$, since
$m_1\sim{t}/{U}$. If we were to retain only the $Um_1n$ term  
and take  $C=1-n$, its value
at $U=\infty$, we would obtain the form of 
Eq.~(\ref{eq:gapequationT=0}) exactly, but
with $m_1$ replacing $n_1$. Thus, from the discussion 
following Eq. (\ref{eq:gapequationT=0}),
superconductivity at large $U$ is driven by a term of 
order ${t^2}/{U}$. This seen clearly in
Fig.~\ref{figure:gap amplitudeII}, where the gap 
amplitude $G$ over the Fermi surface  is
plotted as a function of occupation
 $\left<N\right>$ for different values of ${U}/{|t|}$. 
The peak value of $G$, corresponding
to optimum doping with the square  Fermi surface as before, 
increases rapidly with decreasing
${U}/{|t|}$. At $U=4|t|$ the peak value is about one tenth 
of the rather constant peak
value (see Fig.~\ref{figure:gap amplitude}) found by the 
factorization procedure.
Qualitatively nothing is changed and the quasiparticle 
bands are just like those of
Fig.~\ref{figure:bandgap}, but with a smaller gap. 
The temperature dependence of the gap
amplitude $g(T)$ is calculated from Eq. (\ref{eq:gapequationIII}) 
and in
Fig.~\ref{figure:GTII} the result for $G(T)$ is shown for $U=4|t|$ 
at optimum
doping. If $|t|=0.5$eV the
 corresponding $T_c$ is $10$K.

As discussed earlier we expect the correct values for the gap 
and $T_c$ to lie somewhere between the values obtained  from 
the two approximations used.
Thus for the physically reasonable parameters $|t|=0.5$eV, 
$U=2$eV, we have shown that $d$-wave superconductivity occurs 
with $T_c$ in the range 
$10-100$K at optimum doping $\delta_{c}=0.14$ .

\section{Conclusions}\label{section:conclusions}

We have systematically applied Roth's decoupling 
method~\cite{roth} to the two dimensional
Hubbard model, investigating both the normal and 
superconducting states. 
Although there are intrinsic defects in the method we believe 
we are able to draw 
some significant conclusions. In the normal state the method 
corresponds to a two-pole
approximation to the one-particle Green's function, so that for 
a given wave-vector 
$\vec k$ and spin $\sigma$ the spectral function consists of 
two delta-function peaks,
the sum of whose weights is unity, corresponding to states in 
the upper and lower 
Hubbard bands. For site occupation $\left<N\right> <1$ only 
the lower band is occupied 
and the absence of any incoherent spectral weight below the 
Fermi level necessarily 
implies a Fermi surface of volume greater than that appropriate 
to a non-interacting system 
or Fermi liquid. Nevertheless we have shown in 
Sec.~\ref{section:normalstate} that 
there is very remarkable agreement between calculated bands 
and those obtained in 
Monte Carlo calculations by Bulut {\it et al}.;\cite{bulut} 
in particular for $U=8|t|$ 
and several values of $\left<N\right>$
 the position of the Fermi level, and hence the nature the 
Fermi surface, is in agreement.
This implies that for $\left<N\right> \gtrsim 0.8$ the 
anomalous Fermi surface volume
is not merely a result of the Roth approximation, but is 
a real effect. This could be caused by the finite
temperature in the Monte Carlo calculations, but can 
possibly also be explained
on the basis of the Edwards-Hertz theory,\cite{EH,marco} 
in which for $U$ larger than
the bandwidth, a transition from a Fermi liquid to a 
non-Fermi liquid with enlarged Fermi
surface volume occurs as $\left<N\right>$  increases 
through a critical
value $N_c$. However the two-pole approximation omits 
the quasiparticle lifetime broadening which,
in the Edwards-Hertz theory occurs even at the Fermi 
level for $\left<N\right> > N_c$.
Our calculated bands also feature almost
dispersionless bands around the $(\pi,0)$ points as is 
observed in recent ARPES experiments
on hole doped cuprate superconductors.
\cite{ARPES Bi2212,ARPES Bi2201,ARPES YBCO124,ARPES 
Y124,ARPES YBCO123}
Another feature our calculations share with 
experiments~\cite{ELS,OptSpec} is the weight transfer from 
the upper to 
the lower band upon doping. For intermediate $U$ this 
transfer is $\vec k$-dependent and has a maximum
at the $(\pi,\pi)$ point.

In Sec.~\ref{section:gap} it is shown how 
$d$-wave superconductivity, but not $p$-wave,
occurs. An important conclusion is that reasonable 
transition temperatures, in the range 
$10-100$K, only occur for $U \lesssim 6|t|$ 
and in a fairly small range of occupation 
$\left<N\right>$ near optimum doping where the 
saddle-point feature of the quasiparticle band
structure at $(\pi,0)$ is placed at the Fermi level.
For $U \lesssim 6|t|$ the system is expected 
to be a Fermi liquid for all
$\left<N\right>$, so that the anomalous volume of the 
Fermi surface is spurious in this
regime. However this defect of the method is rather 
fortunate since it enables us to obtain
the saddle-point feature of the band at the Fermi 
level for a doped case ($\left<N\right> <
1$) without recourse to next nearest neighbour hopping. 
Also the absence of lifetime
broadening in the two-pole approximation is a  correct 
feature near the Fermi level in the
Fermi liquid regime. We therefore believe our results 
are significant and closely parallel 
the Monte Carlo results of Husslein {\it et al}.~\cite{husslein} 
on $d$-wave superconductivity in
the $tt'$ Hubbard model. The association of optimum doping 
with a saddle-point at the Fermi
level (see also Ref.\onlinecite{newns}) is consistent with 
the generically different behaviour of
doped superconducting cuprates in many normal state properties 
at exactly optimum doping, as
opposed to those at under- or overdoping.
It is well known~\cite{newns} that a saddle-point at the 
Fermi level leads to  marginal
Fermi liquid transport properties such as a normal state 
resistivity varying linearly with
temperature. A recent discussion of 
thermopower~\cite{newnsthermopower} strengthens the case
for the van Hove scenario in the cuprate superconductors.
 Even though the method of Sec.~\ref{subsection:largeUdecoupling} 
still gives 
superconductivity near optimum doping for $U>8|t|$, 
but with very low $T_c$, it is unlikely
that  the superconducting state would survive the 
introduction of finite quasiparticle
lifetimes at  the Fermi level which might arise 
from non-Fermi liquid behaviour in this
region.

The present work suggests that high temperature 
superconductivity arises in moderately
correlated systems which may be doped so that the 
Fermi level approaches a van Hove
singularity in the quasiparticle band structure.
Correlation helps by flattening the bands, 
accentuating the singularity, but must not be
allowed to reduce site double occupation too markedly. 
Nearest neighbour antiferromagnetic
correlations are particularly important both in 
flattening the bands and in promoting double
occupation. The correlations important for 
superconductivity in the present model are local
ones and the mechanism is different from 
spin-fluctuation-exchange models.

In future work we propose to explore the 
possibility of $s$-wave pairing,
particularly its degree  of anisotropy if it is 
found to occur. Work is in progress on
extending the Roth method consistently to two-particle 
Green's functions, particularly with
the aim of calculating the dynamical susceptibility 
$\chi(\vec{q},\omega)$ and  investigating the
spin dynamics of the system.

\acknowledgements
One of us (J.B.) thankfully acknowledges financial 
support from the SERC, the `Universiteitsfonds Twente', 
the British Council and 
the Commission of the European Community during 
the course of this work.

\newpage

\newpage

\newpage

\newpage

\newpage

\newpage

\newpage

\newpage

\end{document}